\documentclass{amsart}
\usepackage{latexsym}
\usepackage{amssymb, epsf, amscd, verbatim}
\newtheorem{theorem}{Theorem}
\newtheorem{lemma}[theorem]{Lemma}
\newtheorem{proposition}[theorem]{Proposition}
\newtheorem{corollary}[theorem]{Corollary}

\theoremstyle{definition}

\newtheorem{example}[theorem]{Example}

\theoremstyle{remark}
\newtheorem{remark}[theorem]{Remark}

\numberwithin{equation}{section}
\numberwithin{theorem}{section}

\newcommand{\sepbox}[1]{\hfill $\box$ \vspace{4mm}}
\newcommand\lbb[1]{\label{#1}}

\def\tt{\otimes}                               
\def\<{\langle}
\def\>{\rangle}

\def\wti{\widetilde}
\def\what{\widehat}
\def\ov{\overline}

\def\d{\partial}

\def\isoto{\xrightarrow{\sim}}                 
\def\st{\; | \;}                               

\def\Cset{\mathbb{C}}       
\def\Zset{\mathbb{Z}}       

\newcommand{\CC}{\mathbb{C}}

\def\al{\alpha}                         
\def\be{\beta}

\def\de{\delta}
\def\De{\Delta}

\def\la{\lambda}

\def\Th{\Theta}

\def\g{{\mathfrak{g}}}      

\def\ss{{\mathfrak{s}}}
\def\gl{{\mathfrak{gl}}}
\def\sl{{\mathfrak{sl}}}
\def\so{{\mathfrak{so}}}

\def\A{{\mathcal{A}}}
\def\L{{\mathcal{L}}}

\def\F{\mathcal{F}}           

\begin{document}

\title{ Classification of finite simple Lie conformal superalgebras}

\author[D.~Fattori]{Davide Fattori}
\address{Dipartimento di Matematica, Universit\'a di Genova,
via Dodecaneso 35, 16146 Genova, Italy}
\email{fattori@math.mit.edu}
\thanks{The first author was partially supported by CNR-GNSAGA. This
research was partially conducted by the author for the Clay
Mathematics Institute.}

\author[V.~G.~Kac]{Victor G.~Kac}
\address{Department of Mathematics, MIT, Cambridge MA 02139, USA}
\email{kac@math.mit.edu}
\thanks{The second author was partially supported  by NSF grant
DMS-9970007}
\date{\today}

\maketitle

\section{Introduction}\lbb{sintro}
In recent years, two-dimensional conformal field theory has attracted
the attention of many mathematicians and physicists.  In \cite{Bo},
Borcherds introduced the notion of a vertex algebra, which encodes the
operator product expansion (OPE) of chiral fields in this theory. The
singular parts of the OPE (or, equivalently, the commutators of
fields) are encoded by a Lie conformal superalgebra (\cite{K2,K3}).
By means of this formalism, computations of OPE's are greatly
simplified.

In the language of the $\la$-bracket, a Lie conformal superalgebra $R$
is a $\CC [\partial]$-module endowed with a $\CC$-linear map
\begin{displaymath}
  R \otimes R \to \CC [\lambda]\otimes R ,\qquad
  a \otimes b \mapsto [a_{\lambda}b],
\end{displaymath}
satisfying the following axioms \cite{DK,K2} $(a,b,c \in R)$:
\begin{align*}
  \begin{array}{ll}
    \text{(sesquilinearity)}\quad &
    [\partial a_{\lambda}b]=-\lambda [a_{\lambda}b], \;\;
    [a_{\lambda}\partial b]=(\partial +\lambda)[a_{\lambda}b],
    \\
\text{(skew-commutativity)}\quad &
    [b_{\lambda}a]=-(-1)^{p(a)p(b)}[a_{-\lambda -\partial}b] ,\\
    \text{(Jacobi identity)}\quad &
    [a_{\lambda} [b_{\mu}c]] =
    [[a_{\lambda}b]_{\lambda +\mu}c]+
    (-1)^{p(a)p(b)}[b_{\mu}[a_{\lambda}c]] .
  \end{array}
\end{align*}
Finite (i.e. finitely generated as a $\Cset[\d]$-module) simple Lie
conformal algebras were classified in \cite{DK} and their
representation theory was further developed in \cite{CK1, BKV}.

On the other hand, Lie conformal superalgebras are closely connected
to the notion of a formal distribution Lie superalgebra $(\g, \F)$,
i.e. a Lie superalgebra $\g$ spanned by the coefficients of a family
$\F$ of mutually local formal distributions. Namely, to a Lie
conformal superalgebra $R$ one canonically associates the maximal
formal distribution Lie superalgebra $Lie\;
R=R[t,t^{-1}]/\wti{\d}R[t,t^{-1}]$ (see Section \ref{subfdlslcs}),
which establishes an equivalence between the category of Lie conformal
superalgebras and the category of equivalence classes of formal
distribution Lie superalgebras obtained as quotients of $Lie\; R$ by
irregular ideals, see \cite{K2}.

In the present paper, we give the classification of finite simple Lie
conformal superalgebras. The main result is the following theorem
(announced in \cite{K6}):

\begin{theorem}
Any finite simple Lie conformal superalgebra $R$ is isomorphic to one of
the Lie conformal superalgebras of the following list
(see Section \ref{subsomeex} for their construction):
\begin{enumerate}
\item $W_N \;\;(N\geq 0)$;
\item $S_{N,a} \;\; (N \geq 2, \;a \in \Cset)$;
\item $\wti{S}_N \;\; (N \;even,\;\; N \geq 2)$;
\item $K_N \; \; (N \geq 0, \;\; N\neq 4)$;
\item $K_4'$;
\item $CK_6$;
\item $Cur\; \ss$, where $\ss$ is a simple finite-dimensional Lie
superalgebra.
\end{enumerate}
\end{theorem}

The general outline of the proof of this theorem is similar to that of
\cite{DK} in the non ``super'' case.  First of all, we extend to Lie
superalgebras the classical Cartan-Guillemin theorem
(\cite{G1}, \cite{B, BB}) which asserts that  any minimal
non-abelian closed ideal  in
a linearly compact Lie algebra $L$ is  of the form
$\Cset[[t_1,\ldots,t_r]]
\what{\tt} \ss$, where $\ss$ is a simple
linearly compact Lie algebra (see Theorem \ref{cgthm} and Corollary
\ref{cgused}).

Secondly, we deduce that the annihilation algebra $\A(R)$ is an
irreducible central extension of $\Cset[[t_1,\ldots,t_r]]\what{\tt}
\ss$, where $\ss$ is simple linearly compact Lie superalgebra (Lemma
\ref{annauc}).  Recall that $\A(R)$ is the completion of the image of
$R[t]$ in $Lie\;R $.  It is linearly compact if $R$ is finite. The
operator $-\frac{\d}{\d t}$ on $R[t,t^{-1}]$ induces a derivation $T$
of $\A(R)$, and the semi-direct product $\Cset T \ltimes
\A(R)$ is called the extended annihilation algebra.

Thirdly, we remark that the growth of $\A(R)$ is smaller or equal than
one, so that we obtain that either $r=1$ and $\ss$ is a simple
finite-dimensional Lie superalgebra or $r=0$ and $\ss$ is a simple
linearly compact Lie superalgebra of growth $1$, and we may use the
classifications of the papers \cite{K4} and
\cite{K5} respectively. This produces a list of all possible
annihilation algebras (Proposition \ref{pcannia}).

The fourth step is the classification (up to conjugacy) of all even
surjective continuous derivations of these candidates for the
annihilation algebras. This leads to the list of all possibilities for
the extended annihilation algebras (Theorem
\ref{pcanniaader}).

Finally, we reconstruct $R$ from the extended annihilation algebra to
obtain the main result (Theorem \ref{clthm}). An immediate corollary
of this result is a classification of all formal distribution Lie
superalgebras (Corollary \ref{clthmfdls}).

Along the way we show in Section \ref{grow} that the growth of a
simple linearly compact Lie superalgebra is independent of its algebra
filtration, and we classify in Section \ref{scenext} all central
extensions of finite simple Lie conformal superalgebras (which are
important for the construction of simple vertex algebras).

\section{Linearly Compact Lie Superalgebras}\lbb{slclsu}

\subsection{The Cartan-Guillemin theorem}\lbb{subcgthm}

Recall that a \emph{vector superspace} is a $\Zset /2 \Zset$ -graded
vector space, $V=V_{\overline0} \oplus V_{\overline1}$.  We denote by
$p(a)=\al$ the \emph{parity} of an homogeneous element $a$, $a \in
V_{\al}, \; \al \in \{
\ov{0}, \ov{1} \}$.  A subspace $U$ of $V$ is by definition
$\Zset/2\Zset$ -graded, i.e. $U=(U\cap V_{\overline{0}})\oplus(U\cap
V_{\overline{1}})$.  All vector superspaces, linear maps and tensor
products are over the field $\Cset$ of complex numbers. Exterior
and symmetric powers of a vector superspace are to be understood in
the super-sense (see \cite{K4}).

A \emph{superalgebra} $A$ is a vector superspace endowed with an
algebra structure such that $A_{\al}A_{\be} \subset A_{\al +\be}$,
with $\al,\be \in \Zset /2 \Zset$.

A \emph{Lie superalgebra} is a superalgebra satisfying
super-anticommutativity and the super-Jacobi identity (see \cite{K4}).

We endow $\Cset$ with the discrete topology.  Let $V=V_{\overline0}
\oplus V_{\overline1}$ be a Hausdorff topological vector superspace.
We will say that $V$ is a \emph{linearly compact} vector superspace if
every family of closed affine linear varieties of $V$ has non empty
intersection whenever every finite subset of the family has non-empty
intersection.  A topological Lie superalgebra $L$ is called linearly
compact if the underlying topological space is linearly compact.

Let $V^{\star}$ be the topological dual of $V$.  Let $U$ be a linearly
compact subspace of $V$.  We denote by $U^{\perp}$ the set of all
continuous linear functionals which annihilate $U$.  We define a
topology on $V^{\star}$ by taking the collection of all sets of the
form $U^{\perp}$ to be a fundamental system of neighborhoods of the
origin.

In the following we list some properties of linearly compact vector
superspaces.

\begin{proposition}[see \cite{G1}]\lbb{plcvs}

\begin{enumerate}

\item If $A$ is a linearly compact subspace in a linearly compact
vector superspace $V$, then $A$ is closed.

\item Direct products and inverse limits of linearly compact
vector superspaces are linearly compact.

\item A subspace of $V$ is open if and only if it is closed and
has finite codimension.

\item A discrete topological vector superspace is linearly compact if
and only if it is finite-dimensional.

\item If $V$ is discrete (resp. linearly compact), then $V^{\star}$
is linearly compact (resp. discrete).

\item If $V$ is discrete or linearly compact, the
canonical linear map $V \to V^{\star \star}$ is an isomorphism.

\item $V$ is linearly compact if and only if
       it is isomorphic to a (topological) product of
       finite-dimensional discrete spaces.

\item  If $V$ is linearly compact, then it is complete.

\item The image of a linearly compact space under a continuous linear map
is linearly compact.

\item (Chevalley's principle) Suppose $F_1 \supset F_2 \supset \ldots $
is a sequence of closed subspaces in a linearly compact vector
superspace $V$, such that $\cap_i F_i = \{0\}$. If $U$ is a
neighborhood of $0$ in $V$, then there exists an integer $i_0$ such
that $F_{i_0} \subset U$. \hfill $\Box$
\end{enumerate}
\end{proposition}

The basic examples of linearly compact spaces are finite-dimensional
vector superspaces with the discrete topology (see Proposition
\ref{plcvs}(4)) and the space of formal power series $V[[t]]$, where
$V$ is a finite-dimensional vector superspace, with the
\emph{formal topology} defined by taking as a fundamental
system of neighborhoods of the origin the set $\{t^jV[[t]]\}_{j\in
\Zset_+}$ (see Proposition \ref{plcvs}(2)). A closely related important
example is the associative linearly compact superalgebra
$\Cset[[t_1,\ldots,t_r]]\tt\wedge(m)$, where $\wedge(m)$ denotes the
Grassman algebra on $m$ anticommuting indeterminates $\xi_1,\ldots,
\xi_m$ and $p(t_i)=\ov{0}$, $p(\xi_i)=\ov{1}$, with the \emph{formal}
topology defined by $\{(t_1,\ldots,t_r)^j\}_{j \in \Zset_+}$.

Let $V,W$ be linearly compact vector superspaces. Let $V^{\star},
W^{\star}$ be their topological duals.  We form the tensor product
$V^{\star}\tt W^{\star}$, endow it with the discrete topology, and
define the \emph{completed tensor product} of $V$ and $W$ to be the
space $(V^{\star}\tt W^{\star})^{\star}$. It is denoted by
$V\widehat{\tt}W$. Note that, if $dim\: V < \infty$, then
$V\widehat{\tt}W =V \tt W$.

A linearly compact Lie superalgebra $L$ is called \emph{simple} if it
contains no non-trivial closed graded ideals. The same proof as in
\cite{G1}, Proposition 4.3, shows that then $L$ has no non-trivial
graded ideals (closed or not). Due to \cite{Sc}, Proposition 2.1, then $L$
has no non-trivial left or right ideals (graded or not) as well.

\begin{lemma}[Schur's Lemma]\lbb{schlma}
For a topological Lie superalgebra $L$ we set:
\begin{equation*}
\De_L= \{ \tau \in (End L)_{\al},\al \in \Zset_2  \st  \tau([x,y])=
(-1)^{p(x)\al}[x,\tau(y)] \;\;
\text{ for any} \; \; x,y \in L\}.
\end{equation*}
If $L$ is simple and linearly compact, then $\Delta_L=\Cset$.
\end{lemma}
\begin{proof}
Let $\tau\in \Delta_L$.  Note that $ker\; \tau$ is an ideal of
$L$. Since $L$ is simple, by the above discussion either $\tau=0$ or
$\tau$ is invertible, i.e. $\Delta_L$ is a skew-field.  Now one can
argue as in \cite{G1}, Proposition 4.4.
\end{proof}

\begin{lemma}\lbb{isoinpro}
Let $H$ be a closed subalgebra of the linearly compact Lie
superalgebra $L$.  Let $V$ be a linearly compact $H$-module.  Endow
the induced $L$-module $U(L)\tt_{U(H)} V^{\star}$ with the discrete
topology, so that $(U(L)\tt_{U(H)} V^{\star})^{\star}$ is a linearly
compact space.  Endow the $L$-module $Hom_{U(H)}(U(L), V)$ with the
finite-open topology.  Then $Hom_{U(H)}(U(L), V)$ is linearly compact
and is homeomorphic to $(U(L)\tt_{U(H)} V^{\star})^{\star}$ as an
$U(L)$-module.
\end{lemma}

\begin{proof}
The proof is the same as in the ``even'' case, see \cite{B},
Proposition 1, and \cite{BB}, Lemma 1.1.
\end{proof}

Let $L$ be a linearly compact Lie superalgebra and let $V$ be a
linearly compact (resp. discrete) $L$-module.  The space $V$ is called
\emph{topologically} (resp. \emph{algebraically}) irreducible if it
contains no non-trivial closed submodules (resp. no non-trivial
submodules). The module $V$ is called
\emph{topologically} (resp. \emph{algebraically}) absolutely irreducible
if it is topologically (resp. algebraically) irreducible and the
commuting ring of $L$ in $Hom_{\Cset}^{\text{Cont}}(V,V)$
(resp. $Hom_{\Cset}(V,V)$) consists only of scalar operators.  Remark
that $V$ is topologically absolutely irreducible iff $V^{\star}$ is
algebraically absolutely irreducible.

Let $I$ be a closed ideal of the linearly compact Lie superalgebra
$L$.  Let $V$ be a topological $I$-module.  The \emph{stabilizer} of
$V$ is defined as follows:
\begin{equation*}
H=\{ x\in L \st \exists s \in Hom_{\Cset}^{\text{Cont}}(V,V) \st
[x,z]v=[s,z]v \;\; \text{for any}\;\; z \in I, \;\;v\in V \}.
\end{equation*}
Then $H$ is a closed subalgebra of $L$ containing $I$.

\begin{theorem}[Blattner] \lbb{irrindrep}
Let $I$ be a closed ideal in a linearly compact Lie superalgebra $L$.
Let $V$ be an algebraically absolutely irreducible discrete
$I$-module, and let $H$ be its stabilizer.  Let $W$ be an
algebraically absolutely irreducible discrete $H$-module such that, as
an $I$-module, it is a direct sum of copies of $V$. Then
$U(L)\tt_{U(H)} W$ is an algebraically absolutely irreducible
$L$-module.
\end{theorem}



\begin{proof}
As in the ``even'' case, see \cite{BB}, Theorem 3(b).
\end{proof}


\begin{proposition}\lbb{topirrindrep}
In the notation of Theorem \ref{irrindrep}, let $V$ be a topologically
absolutely irreducible $I$-module.  Let $W$ be a topologically
irreducible linearly compact $H$-module.  Suppose that, as an
$I$-module, it is topologically module-isomorphic to a direct product
of copies of $V$. Then the $L$-module $Hom_{U(H)}(U(L),W)$ is
topologically absolutely irreducible.
\end{proposition}
\begin{proof}
As in the ``even'' case, see \cite{B}, Theorem 1.2.
\end{proof}

\begin{proposition}\lbb{notcg}
Let $L$ be a linearly compact Lie superalgebra.  Suppose $L$ admits a
non-abelian minimal closed ideal $I$.  Then $I$ possesses a maximal
proper ideal $J$, which is closed, $\ov{I}:=I/J$ is a simple
non-abelian linearly compact Lie superalgebra and $N:=N_L(J)$ is
open. Let $\varphi$ be the canonical map of $I$ onto $\ov{I}$. Then
$\ov{I}$ is a $N$-module and $\varphi$ is a
$N$-homomorphism. Furthermore, we have a $L$-module homomorphism
\begin{equation*}
\Th: I \to Hom_{U(N)}(U(L),\ov{I}),
\end{equation*}
where $\Th(x)(a)=(-1)^{p(a)p(x)}\varphi((ad \; a)(x))\;\; (x\in I,\;
a\in U(L))$.
\end{proposition}
\begin{proof}
As in the ``even'' case, see \cite{B}, Lemma 2.2.
\end{proof}


\begin{theorem}[Cartan-Guillemin]\lbb{cgthm}
Let $I$ be a non-abelian minimal closed ideal in a linearly compact
Lie superalgebra $L$.  Then $I$ is homeomorphic via $\Th$ to
$Hom_{U(N)}(U(L),\ov{I})$ both as a Lie superalgebra and as a
$L$-module.
\end{theorem}


\begin{proof}
We can apply the Schur Lemma \ref{schlma} to $\ov{I}$ and Proposition
\ref{topirrindrep} and \ref{notcg} in order to use the same argument
as in \cite{B}, Theorem 2.4.
\end{proof}


\begin{corollary}\lbb{cgused}
Let $dim(L/N)_{\ov{0}}=r$ and $dim(L/N)_{\ov{1}}=m$. Then, in the
notation of Theorem \ref{cgthm},
\begin{equation*}
I \isoto (\Cset[[ t_1, \ldots, t_r ]] \tt
\wedge(m)) \what{\tt} \ov{I}
\end{equation*}
as topological Lie superalgebras.
\end{corollary}

\begin{proof}
By \cite{B}, Corollary to Proposition 7, $Hom_{U(N)}(U(L),\ov{I})$ is
isomorphic to $Hom_{\Cset}(S(L/N),\ov{I})$ as a topological Lie
superalgebra.  Since $N$ is open, $dim(L/N) < \infty$. The fact that a
linearly compact space can be identified with its double dual implies
that $Hom_{\Cset}(S(L/N),\ov{I})$ is isomorphic to $(\Cset[[ t_1,
\ldots, t_r ]] \tt \wedge(m)) \widehat{\tt} \ov{I}$.
\end{proof}

\subsection{Growth}\lbb{grow}

A \emph{filtration} of a vector superspace $V$ is a decreasing
filtration by subspaces of finite codimension ($j_0 \in \Zset$):
\begin{displaymath}
V=V_{j_0} \supset V_{j_0+1} \supset V_{j_0+2} \supset \ldots
\end{displaymath}
such that $\cap_j V_j =\{ 0 \}$.  The \emph{growth} of this filtration
is defined as follows:
\begin{displaymath}
gw(V)=\limsup_{j\to \infty}\frac{\log \;dim(V/V_j)}{\log \;j}.
\end{displaymath}
Given a subspace $U$ of $V$, one has an induced filtration on $U$ and
on $V/U$. Clearly, $gw(U)\leq gw(V)$ and $gw(V/U)\leq gw(V)$.  Also,
if the filtration $\{ V_j \}$ of $V$ is shifted (i.e. we take
$V_{(j)}=V_{j+a}$) or rescaled (i.e. we take $V_{(j)}=V_{nj}$ for a
fixed positive integer $n$) then $gw(V)$ remains unchanged.  The first
claim is obvious, while the second one follows from the observation
that on the one hand, the rescaling of the filtration may obviously
only increase the growth, but, on the other hand, it may only decrease
the growth:
\begin{displaymath}
gw(V) \geq \limsup_{j\to \infty}\frac{\log \;dim(V/V_{nj})}{\log \;nj}
= \limsup_{j\to \infty}\frac{\log \;dim(V/V_{nj})}{\log \;j}.
\end{displaymath}

An \emph{algebra filtration} of a linearly compact Lie superalgebra
$L$ is a filtration of $L$ by open subspaces $L_i$ such that
$[L_i,L_j]\subseteq L_{i+j}$. A similar definition applies to
associative algebras.

Recall that a \emph{fundamental subalgebra} of a linearly compact Lie
superalgebra $L$ is a proper open subalgebra that contains no non-zero
ideals of $L$. Given a fundamental subalgebra $L_0$ of $L$ one
constructs the \emph{canonical} (algebra) filtration of $L$ associated
to $L_0$:
\begin{displaymath}
L=L_{-1} \supset L_0 \supset L_1 \supset \ldots
\end{displaymath}
by letting inductively for $j \geq 1$ (cf. \cite{GS}):
\begin{displaymath}
L_j=\{ a \in L_{j-1} \st [a,L] \subset L_{j-1} \}.
\end{displaymath}
One knows \cite{G1} that any linearly compact Lie superalgebra $L$
contains a proper open subalgebra $L_0$, which is of course
fundamental if $L$ is simple.

One defines $gw(L,L_0)$ to be the growth of the canonical filtration
of $L$ associated to $L_0$.

\begin{proposition}[\cite{BDK}]\lbb{gwwd}
The number $gw(L,L_0)$ is independent of the choice of the fundamental
subalgebra $L_0$ of $L$.
\end{proposition}

\begin{proof}
If we choose the $j$-th member $L_j$ ($j \geq1$) of the canonical
filtration associated to $L_0$ as another fundamental subalgebra, say
$\wti{L_0}$, then, by definition, the associated canonical filtration
is $\wti{L_k}=L_{k+j}$. Hence this change of the fundamental
subalgebra does not affect the growth.

Now, if $M_0$ is another fundamental subalgebra of $L$ and $\{M_j\}$
is the associated canonical filtration, then, by Chevalley's
principle, $L_k \subset M_0$ for sufficiently large $k$, hence
$L_{k+j} \subset M_j$ for all $j$. Therefore, $gr(L,L_0) \geq
gr(L,M_0)$. Exchanging the roles of $L_0$ and $M_0$ we get the
opposite inequality.
\end{proof}

We denote by $gw(L)$ and call the \emph{growth} of $L$ the number
$gw(L,L_0)$ defined above.  If $L$ is simple, one can prove the
following stronger result.

\begin{theorem}\lbb{grsilc}
Let $L$ be a simple linearly compact Lie superalgebra. Then any
algebra filtration of $L$ by open subspaces has growth equal to
$gw(L)$.
\end{theorem}

\begin{proof}
If $L$ contains an open subalgebra $M_1$ such that all $M_j=M_1^j$ are
open subspaces of $L$ (then this is automatically an algebra
filtration), let us denote by $gw'(L,M_1)$ the growth of this
filtration. Since the growth is invariant under rescaling, we see
(using Chevalley's principle as in the proof of Proposition
\ref{gwwd}) that $gw'(L,M_1)=gw'(L)$ is independent of the choice of
$M_1$. Furthermore, we clearly have:
\begin{equation}\lbb{grine}
gw(L)\leq gw(\text{any algebra filtration}) \leq gw'(L),
\end{equation}
if we choose $M_1$ to be a fundamental subalgebra of $L$.

Furthermore, it follows from the classification of simple linearly
compact Lie superalgebras \cite{K5} that $L$ has a Weisfeiler
filtration
\begin{equation*}\lbb{wesfi}
L=L_{-d}^W \supset L_{-d+1}^W \supset \ldots \supset L_{-1}^W \supset
L_0^W
\supset L_1^W \supset \ldots
\end{equation*}
for which the associated graded Lie superalgebra $Gr(L)=\oplus_{j\geq
-d}\: \g_j$ has the property that the space $\g_1 + \g_2$ generates
$\oplus_{j\geq 1} \g_j$ (in fact, in all cases except for
$L=Der(\Cset[[t]])$, one has $\g_j=\g_1^j$, $j >0$), hence
\begin{equation}\lbb{othfil}
L_{2k}^W \subset (L_1^W)^k \subset L_k^W \quad \text{for all } \;\; k.
\end{equation}
On the other hand, for the canonical filtration $\{ L_j \}$ associated
to $L_0^W$ one has
\begin{equation}\lbb{omineq}
L_j \subset L_{jd}^W \quad \text{for all} \;\; j.
\end{equation}
It follows from (\ref{othfil}) and (\ref{omineq}) that $gw(L)\geq
gw'(L)$. Therefore, by (\ref{grine}), the growth of any algebra
filtration of $L$ is equal to $gw(L)$.
\end{proof}

The following theorem follows from the classification of simple
linearly compact Lie superalgebras \cite{K5}.

\begin{theorem}\lbb{gr1lcls}
Any simple linearly compact Lie superalgebra of growth at most $1$ is
either finite-dimensional or is isomorphic to one of the following Lie
superalgebras (see Section \ref{subsomeex} for their description):
$\ov{W}(1,N)$, $N\geq 0$; $\ov{S}(1,N)'$, $N\geq 2$; $\ov{K}(1,N)$,
$N\geq 0$; $\ov{E}(1,6)$. \hfill $\Box$
\end{theorem}

\subsection{On derivations of linearly compact Lie superalgebras}

In this Section we prove three propositions which will be essential in
the sequel. We shall denote by $Der(L)$ the Lie superalgebra of all
continuous derivations of a topological superalgebra $L$.

\begin{proposition}\lbb{detp}
Let $A$ be a commutative, associative, unital linearly compact
superalgebra and let $L$ be a simple linearly compact Lie
superalgebra.  Then
\begin{equation*}
Der(A \what{\tt}L)= Der(A) \tt 1 + A\what{ \tt} Der(L).
\end{equation*}
\end{proposition}

\begin{proof}
Using Schur Lemma (see Proposition \ref{schlma}), one can apply the
same argument as in \cite{BDK}, Proposition 6.12.
\end{proof}

\begin{proposition}\lbb{derconj}
Let $L$ be a linearly compact Lie superalgebra and let
\begin{equation*}
L=L_{-d} \supset L_{-d+1} \supset \ldots \supset L_0
\supset L_1 \supset \dots
\end{equation*}
be an algebra filtration of $L$ of depth $d>0$. Let $D$ be an even
element of $L$ such that
\begin{equation}\lbb{derfilpr}
[D,L_k]= L_{k-d} \;\;\; \text{for all} \;\; k \geq d.
\end{equation}
Let $V$ be a finite-dimensional Lie algebra acting by derivations on
$L$ such that
\begin{equation}\lbb{outderfilpr}
V(L_k) \subseteq L_k \quad \text{for all} \quad k \geq 0,
\end{equation}
and let $\wti{L}=V \ltimes L$.  Then any even element of the form
$D+v+g_0 \in \wti{L}$, where $v \in V$ and $g_0 \in L_0$, can be
conjugated via a continuous inner automorphism of $L$ to $D+v$.

\end{proposition}

\begin{proof}
Let $m$ be the maximal integer such that $g_0 \in L_m \backslash
L_{m+1}$.  Then there exists $l_{m+d} \in L_{m+d}$ such that
$[D,l_{m+d}]=g_0$. The automorphism $exp(ad (l_{m+d}))$ is
well-defined and converges uniformly on $\wti{L}$, and we have:
\begin{align*}
exp(ad (l_{m+d}))(D+v+g_0)&= D +v +(g_0 + [l_{m+d},D]) + \\
&+ ([l_{m+d},g_0]-v(l_{m+d})) + \ldots  \\
&=D+v +\text{terms in} \;\;L_{m+d}.
\end{align*}
By repeating this argument, we obtain $D+v$ in the limit.
\end{proof}

\begin{proposition}\lbb{dernonsofd}
Any non-solvable finite-dimensional Lie superalgebra $\g$ has no even
surjective derivations.
\end{proposition}
\begin{proof}
Let $D$ be an even surjective derivation of $\g$. It clearly
transforms $\g_{\ov{0}}$ surjectively into itself. Moreover, $D$
leaves the radical $\mathfrak r$ of $\g_{\ov{0}}$ invariant, hence it
induces a derivation of $\g_{\ov{0}}/\mathfrak r$ which is not inner
because it is surjective. On the other hand, $\g_{\ov{0}}/\mathfrak r$
is a semisimple Lie algebra, so that every derivation is
inner. Consequently, $\g_{\ov{0}}=\mathfrak r$ is solvable, but this
in turn implies (see
\cite{K4}, Proposition 1.3.3) that $\g$ is solvable.
\end{proof}

\section{Formal Distribution Lie Superalgebras
and Lie Conformal Superalgebras}
\lbb{subfdlslcs}
\subsection{Basic definitions}

Let $V$ be a vector superspace. A $V$-valued \emph{formal
distribution} in one indeterminate $z$ is a formal power series
\begin{equation*}
a(z)=\sum_{n \in \Zset}a_{n}z^{-n-1}, \quad a_n \in V.
\end{equation*}

The vector superspace of all formal distributions in one indeterminate
will be denoted by $V[[z,z^{-1}]]$.  It has a natural structure of a
$\Cset[\d_z]$-module. We define
\begin{equation*}
Res_z \; a(z) = a_{-1}.
\end{equation*}
Similarly, one can define a formal distribution in two indeterminates:
\begin{equation*}
a(z,w)=\sum_{m,n \in \Zset}a_{m,n}z^{-m-1}w^{-n-1}.
\end{equation*}
It is called \emph{local} if
\begin{equation*}
 (z-w)^Na(z,w)=0 \quad \text{for} \quad N \gg 0.
\end{equation*}

Let $\g$ be a Lie superalgebra, and let $a(z),b(z)$ be two $\g$-valued
formal distributions. They are called local if $[a(z),b(w)]$ is local,
i.e.
\begin{equation*}
(z-w)^N [a(z),b(w)]=0 \quad \text{for} \quad N \gg 0.
\end{equation*}

Let $\g$ be a Lie superalgebra, and let $\F$ be a family of
$\g$-valued mutually local formal distributions. The pair $(\g,\F)$ is
called a \emph{formal distribution Lie superalgebra} if $\g$ is
spanned by the coefficients of all formal distributions from $\F$.

\begin{proposition}[\cite{K2}]
Two $\g$-valued formal distributions $a(z),b(z)$ are local iff
\begin{equation}\lbb{locfd}
[a(z),b(w)]=\sum_{j \in \Zset_+}c^j(w) \d_w^j \de(z-w)/j!,
\end{equation}
where the sum is finite. Here $\de(z-w)=\sum_{n\in \Zset}z^n w^{-n-1}$
is the formal delta-function, and the OPE coefficients $c^j(w) \in
\g[[w,w^{-1}]]$ can be computed as follows:
\begin{equation}\lbb{opecoe}
c^j(w)=Res_z(z-w)^j[a(z),b(w)].
\end{equation}
\end{proposition}

The algebraic analogue of the Fourier transform, see \cite{K2},
provides an effective way to study the OPE.

The \emph{formal Fourier transform} of a formal distribution in two
indeterminates is defined as follows:
\begin{equation*}
F^{\la}_{z,w}(a(z,w))=Res_z e^{\la (z-w)}a(z,w).
\end{equation*}
One has:
\begin{equation*}
F^{\la}_{z,w}(\d^j_w \de(z-w))=\la ^j,
\end{equation*}
therefore the formal Fourier transform is the generating series of the
OPE coefficients of $[a(z),b(w)]$.  The $\la$\emph{-bracket} of $a(w)$
and $b(w)$ is defined as
\begin{equation*}
[a(w)_{\la}b(w)]=F^{\la}_{z,w}([a(z),b(w)]).
\end{equation*}
The coefficient of $\la^n/n!$ in the $\la$-bracket is a $\g$-valued
formal distribution, called the $n$-th product of $a(w)$ and $b(w)$
(it is computed by formula (\ref{opecoe})).

The properties of the $\la$-bracket lead to the following basic
definition (see \cite{K2,K3}).


A \emph{Lie conformal superalgebra} $R$ is a left
$\Zset/2\Zset$-graded $\Cset[\d]$-module endowed with a $\CC$-linear
map, called the $\la$\emph{-bracket},
\begin{displaymath}
  R \otimes R \to \CC [\lambda]\otimes R ,\qquad a \otimes b \mapsto
  [a_{\lambda}b]
\end{displaymath}
satisfying the following axioms $(a,b,c \in R)$:
\begin{align*}
\begin{array}{ll}
    \text{(sesquilinearity)}\quad &
    [\partial a_{\lambda}b]=-\lambda [a_{\lambda}b], \;\;
    [a_{\lambda}\partial b]=(\partial +\lambda)[a_{\lambda}b],
    \\
    \text{(skew-commutativity)}\quad &
    [b_{\lambda}a]=-(-1)^{p(a)p(b)}[a_{-\lambda -\partial}b] ,\\
\text{(Jacobi identity)}\quad &
    [a_{\lambda} [b_{\mu}c]] =
    [[a_{\lambda}b]_{\lambda +\mu}c]+ (-1)^{p(a)p(b)}
   [b_{\mu}[a_{\lambda}c]] .
\end{array}
\end{align*}
We write
\begin{equation}
[a_{\lambda}b]=\sum_{n \in \Zset_+}\frac{\la^n}{n!}(a_{(n)}b).
\end{equation}
The coefficient $a_{(n)}b$ is called the $n$-\emph{th product} of $a$
and $b$.  A \emph{subalgebra} $S$ of $R$ is a $\Cset[\d]$-submodule of
$R$ such that $S_{(n)}S \subseteq S$ for any $n \in \Zset_+$.  An
\emph{ideal} $I$ of $R$ is a $\Cset[\d]$-submodule of $R$ such that
$R_{(n)}I \subseteq I$ for any $n \in \Zset_+$.  A Lie conformal
superalgebra $R$ is \emph{simple} if it has no non-trivial ideals and
the $\la$-bracket is not identically zero. A Lie conformal
superalgebra $R$ is \emph{finite} if it is finitely generated as a
$\Cset[\d]$-module. We denote by $R'$ the derived subalgebra of $R$,
i.e. the $\Cset$-span of all $n$-th products.

One knows that any torsion element of $R$ (viewed as a
$\Cset[\d]$-module) has zero $\la$-bracket with $R$
\cite{DK}, hence  a finite simple Lie conformal superalgebra is
free as a $\Cset[\d]$-module.

One can associate to a formal distribution Lie superalgebra $(\g, \F)$
a Lie conformal superalgebra as follows. Let $\ov{\F}$ be the minimal
$\Cset[\d_z]$-submodule of $\g[[z,z^{-1}]]$ that contains $\F$ and is
closed under all $n$-th products, $n \in \Zset_+$. Then the
$\la$-bracket defines a Lie conformal superalgebra structure on
$\ov{\F}$.

Vice versa, given a Lie conformal superalgebra $R$, we can construct a
formal distribution Lie superalgebra $Lie\; R$ using the following
definition.

The \emph{affinization} of a Lie conformal superalgebra $R$ is the Lie
conformal superalgebra
\begin{equation*}
\wti{R}= R \tt \Cset[t,t^{-1}],
\end{equation*}
with $p(t)=\ov{0}$, the $\d$-action defined by $\wti{\d}=\d \tt 1 +
1 \tt \d_t$ and the $n$-th products defined by:
\begin{equation*}
(a\tt f)_{(n)}(b \tt g) = \sum_{j \geq 0} (a_{(n+j)}b) \tt
((\d_t^{j}f)g)/j!,
\end{equation*}
where $a,b \in R, \; f,g \in \Cset[t,t^{-1}]$ and $n \in \Zset_+$.

We let $Lie\;R= \wti{R}/\wti{\d}\wti{R}$ and denote by $a_n$ the image
of $a \tt t^n$ in $ Lie\;R$. Then $ Lie\;R$ is a Lie superalgebra with
respect to the $0$-th product induced from $\wti{R}$. Explicitly, the
bracket is
\begin{equation}\lbb{brinannia}
[a_m,b_n]=\sum_{j \geq 0}
\binom{m}{j}
(a_{(j)}b)_{m+n-j},
\quad a,b \in R;\;\;  m,n \in \Zset.
\end{equation}
Also,
\begin{equation}
(\d a)_m=-m a_{m-1}, \quad a \in R, \quad m \in \Zset.
\end{equation}
The Lie superalgebra $Lie \;R$ admits an even derivation $T$ defined
as $Ta_{n}=-na_{n-1}$.

Let
\begin{equation*}
\F(R)=\{ a(z)=\sum_{n \in \Zset  } a_n z^{-n-1}, \; a\in R \}.
\end{equation*}
Then (\ref{brinannia}) is equivalent to (\ref{locfd}), where
$c^j(w)=(a_{(j)}b)(w)$, hence all formal distributions in $\F(R)$ are
pairwise local.  The pair $(Lie\;R, \F(R))$ is called \emph{ the
maximal formal distribution Lie superalgebra} associated to the Lie
conformal superalgebra $R$.

Let $(\g, \F)$ be a formal distribution Lie superalgebra.  An ideal in
$\g$ is called \emph{irregular} if it does not contain all the
coefficients of a non-zero element of $\ov{\F}$. Then any formal
distribution Lie superalgebra $(\g, \F)$ such that $\ov{\F}\simeq R$
is a quotient of $Lie\;R$ by an irregular ideal \cite{K2}.

An ideal $I$ in $(\g, \F)$ is called \emph{regular} if it is of the
form $I=\{ a_n \st a\in J, \; n \in \Zset_+ \}$, where $J$ is an ideal
of the Lie conformal superalgebra $\ov{\F}$; $I$ is clearly
$T$-stable.

Let $(Lie\;R, \F(R))$ be the maximal formal distribution Lie
superalgebra associated to the Lie conformal superalgebra $R$.  We let
\begin{equation*}
 (Lie\;R)_- = \< \; a_n \st a\in R, \; n \in \Zset_+ \; \>, \;\;
 (Lie\;R)_+= \< a_n \st a\in R, \; n <0  \>.
\end{equation*}
Formula (\ref{brinannia}) implies that these are both $T$-invariant
subalgebras of $Lie\;R$.

Let $R$ be a finite Lie conformal superalgebra, and let $\{a^j\}_{j
\in J}$ be a finite set of generators of $R$ as a $\Cset[\d]$-module.
Let $\L_m$ be the $\Cset$-span of $\{ a^j_i \st i \geq m, \; j \in J
\}$.

It is easy to see using (\ref{brinannia}) that the subspaces $\L_m$
form a
\emph{quasi-filtration} of $(Lie\;R)_-$ (see \cite{DK}):
\begin{equation}\lbb{filanna}
(Lie\;R)_-=\L_0 \supset \L_1 \supset \L_2 \supset \ldots
\end{equation}
by subspaces of finite codimension, i.e. $\cap_i \L_i= \{0\}$ and
there exists an integer $d\in \Zset_+$ such that
\begin{equation}\lbb{qfilanna}
[\L_i,\L_j] \subseteq  \L_{i+j-d},  \quad  i,j \in \Zset_+.
\end{equation}
It also follows from the construction that
\begin{equation}\lbb{dimqufi}
dim\L_i/\L_{i+1} \leq |J|, \quad i \in \Zset_+.
\end{equation}
Furthermore,
\begin{equation}\lbb{conoft}
[T,\L_i]= \;\L_{i-1}.
\end{equation}
Due to Chevalley's principle, the completion $\A(R)$ of $(Lie\; R)_-$
with respect to the topology induced by this filtration is independent
of the choice of the $\Cset[\d]$-generators of $R$.  The Lie
superalgebra $\A(R)$ is a linearly compact Lie superalgebra, called
the
\emph{ annihilation algebra} of $R$. Note that formula
(\ref{qfilanna}) (resp. (\ref{conoft})) implies that the bracket
(resp. $T$) is continuous on $\A(R)$.  The map $T$ is surjective on
$(Lie\; R)_-$ by its very definition, hence it extends to an even
continuous surjective derivation of $\A(R)$ (because $\A(R)$ is a
Hausdorff topological space). The (linearly compact) Lie superalgebra
\begin{equation*}
\A(R)^e=\Cset T \ltimes \A(R)
\end{equation*}
is called the \emph{extended annihilation algebra} of $R$.

\begin{proposition}\lbb{companna}
Let $(\g, \F)$ be a formal distribution Lie superalgebra. Suppose that
the associated Lie conformal superalgebra $\ov{\F}$ is free as a
$\Cset[\d]$-module and $rank \; \ov{\F}= \vert \F \vert < \infty$.
Also, assume that all the coefficients of the formal distributions in
$\F$ form a $\Cset$-basis of $\g$. Then $(\g, \F)$ is the maximal
formal distribution Lie superalgebra associated to $\ov{\F}$.
\end{proposition}

\begin{proof}
Let $\F= \{ a^i(z), \; i=1,\ldots, N \}$. Then $\{ a^i, \; i=1,\ldots,
N \}$ is a $\Cset[\d]$-basis of the free $\Cset[\d]$-module $\ov{\F}$
and $Lie \;\ov{\F}$ is spanned by the set $\{ a^i_n, \;i=1,\ldots,N,
\;\; n\in \Zset \}$. We have a canonical
surjective map $Lie \;\ov{\F} \to \g$, and since the images of the
$a^i_n$'s are linearly independent in $\g$, the kernel of this map is zero.
\end{proof}

\subsection{Some examples}\lbb{subsomeex}

\begin{example}\lbb{loop}
Let $\g$ be a finite-dimensional Lie superalgebra. The \emph{loop
algebra} associated to $\g$ is the Lie superalgebra
\begin{equation*}
\wti{\g}= \g[t,t^{-1}], \;\;
p(at^k)=p(a) \;\; \text{for} \;\; a\in \g,\; k\in \Zset,
\end{equation*}
with bracket
\begin{equation*}
[a\tt t^n,b\tt t^m]=[a,b]\tt t^{n+m} \quad (a,b \in \g; \;m,n \in
\Zset).
\end{equation*}
We introduce the family $\F_{\g}$ of formal distributions (known as
currents)
\begin{equation*}
a(z)=\sum_{n \in \Zset} (a\tt t^n) \;z^{-n-1}, \quad a\in \g.
\end{equation*}
It is easily verified that
\begin{equation*}
[a(z),b(w)]= [a,b](w)\de(z-w),
\end{equation*}
hence $(\wti{\g}, \F_{\g})$ is a formal distribution Lie superalgebra.
The associated Lie conformal superalgebra is $\Cset[\d] \tt \g$, with
$\la$-bracket (we identify $1 \tt \g$ with $\g$)
\begin{equation*}
[a_{\la}b]=[a,b], \quad a,b \in \g;
\end{equation*}
it is called the \emph{current conformal algebra} associated to $\g$,
and it is denoted by $Cur\; \g$.  The Lie conformal superalgebra
$Cur\; \g$ is simple iff $\g$ is a simple Lie superalgebra. Indeed, an
ideal $I$ of $\g$ generates the ideal $\Cset[\d]\tt I$ of $Cur \;
\g$. Conversely, let $J=\Cset[\d]\tt V$ be an ideal of $Cur \;
\g$. Take a non-zero $a=\sum_i \d^ia_i \in J$, where $a_i \in
\g$. Then for any $b \in \g$ we have:
$[a_{\la}b]=\sum_i\frac{(-\la)^i}{i!}(a_{(i)}b)$, hence $a_{(i)}b \in
V$ for all $i \geq 0$, and $V$ is an ideal in $\g$. Due to Proposition
\ref{companna}, $(\wti{\g}, \ov{\F_{\g}})$ is the maximal formal
distribution Lie superalgebra $(Lie\; Cur\; \g, \F(Cur\; \g))$
associated to $Cur
\;\g$; also, it is clear that $T=-\frac{\d}{\d t}$. Hence the
annihilation algebra and the extended annihilation algebra are
respectively $\g[[t]]$ and $\Cset\frac{\d}{\d t} \ltimes
\g[[t]]$.

\end{example}

\begin{example}\lbb{wser}
Let us denote by $\ov{\wedge}(1,N)$ (resp. $\wedge((1,N))$) the tensor
product of $\Cset[[x]]$ (resp. $\Cset[x,x^{-1}]$) and the exterior
algebra $\wedge(N)$ in the indeterminates $\xi_1,\ldots,
\xi_N$.  They are associative, commutative superalgebras if we set
$p(x)=\ov{0},\; p(\xi_i)=\ov{1}, \; i=1, \ldots, N$ and
$\ov{\wedge}(1,N)$ is a linearly compact algebra in the formal
topology. Let $\ov{W}(1,N)$ (resp. $W((1,N))$) be the Lie superalgebra
of all continuous derivations of $\ov{\wedge}(1,N)$ (resp. all
derivations of $\wedge((1,N))$). Then $\ov{W}(1,N)$ is a simple
linearly compact Lie superalgebra \cite{K5}.  Occasionally we will be
dealing also with the Lie superalgebra $W(1,N)$ of all derivations of
$\wedge(1,N)=\Cset[\d]\tt\wedge(N)$ and its subalgebras, but the main
role will be played by their completions in the formal topology.
Every element of $\ov{W}(1,N)$ (resp. $W((1,N))$) can be written in
the form
\begin{equation}\lbb{eq:2.1}
D=\sum_{i=0}^N P_i \d_i,
\end{equation}
where $P_i \in \ov{\wedge}(1,N)$ (resp. $\wedge((1,N))$);
$\d_0:=\frac{\d}{\d x}$ is an even derivation and $\d_i:=\frac{\d}{\d
\xi_i}, \;\; i=1, \ldots, N,$ are odd derivations.

For each element $A \in \wedge(N)$, and for any $j=0,1, \ldots ,N$ we
define a $W((1,N))$-valued formal distribution
\begin{equation*}
A^j(z)=\sum_{n \in \Zset} (x^n  A) \d_j \; z^{-n-1}.
\end{equation*}
The commutation relations are ($A,B \in \wedge(N) \;\; \text{and}
\;\;i,j=1,\ldots ,N$):
\begin{align*}
[A^i(z),B^j(w)]&=((A\d_iB)^j(w) + (-1)^{p(A)}((\d_jA)B)^i(w))\de(z-w);
\\ [A^i(z), B^0(w)]&=(A\d_iB)^0(w)\de(z-w) -(-1)^{p(B)}(AB)^i(w)\d_w
\de(z-w); \\ [A^0(z),B^0(w)]&=-\d_w(AB)^0(w) \de(z-w) -2(AB)^0(w)\d_w
\de(z-w).
\end{align*}
Hence the family $\F_W=\{A^j(z)\}_{A \in \wedge(N), \; j=0,\ldots,N}$
consists of mutually local formal distributions and $(W((1,N)),\F_W)$
is a formal distribution Lie superalgebra.

The associated Lie conformal superalgebra is
\begin{equation*}
W_N=\Cset[\d]\tt \left( W(N) \oplus \wedge(N) \right),
\end{equation*}
where $W(N)$ denotes the Lie superalgebra of all derivations of $\wedge(N)$.
The  $\la$-bracket $(a,b \in W(N); f,g \in \wedge(N))$ is as follows:
\begin{equation}\lbb{lbrwn}
[a_{\la}b]=[a,b],\;\;\; [a_{\la}f]=a(f)-\la (-1)^{p(a)p(f)}fa, \;\;\;
[f_{\la}g]=-\d(fg) -2\la fg.
\end{equation}
The Lie conformal superalgebra $W_N$ is simple for $N \geq 0$. Indeed,
it is easy to see that $W_0$ and $W_1$ are simple. Let $I$ be an ideal
of $W_N$.  Taking $[1_{\la}I]$ we conclude from (\ref{lbrwn}) that $I$
equals to the sum of its intersections with $Cur \; W(N)$ and
$\Cset[\d]\tt \wedge(N)$.  If $N\geq 2$, then $Cur \; W(N)$ is simple,
hence either $I\subset \Cset[\d]\tt \wedge(N)$ or $I\subset Cur \;
W(N)$. Formula (\ref{lbrwn}) implies that the first case is impossible
and that in the second case $I=W_N$.  Furthermore, $W_N$ is a free
$\Cset[\d]$-module of rank $(N+1)2^N$.

We shall need the following representation of $W_N$ on $\Cset[\d]\tt
\wedge(N)$ (see \cite{CK1}):
\begin{equation}\lbb{lactwn}
a_{\la}g=a(g), \quad
f_{\la}g=-(\d + \la)fg, \quad  a\in W(N); f,g \in \wedge(N).
\end{equation}

By Proposition \ref{companna},
\begin{equation*}
\A(W_N)=\ov{W}(1,N) \quad \text{and} \quad
\A(W_N)^e=\Cset ad(\d_0) \ltimes \A(W_N).
\end{equation*}

\end{example}


\begin{example}\lbb{sser}
Recall that the \emph{divergence} of a differential operator
(\ref{eq:2.1}) is defined by the formula
\begin{equation*}
div\;D= \d_0 P_0 + \sum_{i=1}^N (-1)^{p(P_i)}\d_i P_i;
\end{equation*}
its main property is
\begin{equation*}
div\;[D_1,D_2]=D_1(div\; D_2)- (-1)^{p(D_1)p(D_2)}D_2(div\; D_1).
\end{equation*}
It follows that
\begin{equation}
S((1,N))\;(\text{resp.}\; \ov{S}(1,N))=\{ D \in W((1,N))
\;(\text{resp.}\; \ov{W}(1,N))\st div\; D=0 \}
\end{equation}
is a subalgebra of the Lie superalgebra $W((1,N))$
(resp. $\ov{W}(1,N)$).  We have
\begin{equation*}
S((1,N))\;(\text{resp.}\; \ov{S}(1,N))=S((1,N))'\;(\text{resp.}\;
\ov{S}(1,N)') \oplus \Cset \xi_1 \ldots \xi_N
\d_0,
\end{equation*}
where $S((1,N))'$ (resp. $\ov{S}(1,N)'$) denotes the derived
subalgebra.  The Lie superalgebra $\ov{S}(1,N)'$ is a simple linearly
compact Lie superalgebra for $N\geq 2$ \cite{K5}.

Let $\{ A_i, \; i=1,\ldots, (N-1)2^N +1 \}$ be a basis of $S(N)$, the
$0$-divergence subalgebra of $W(N)$, and let $\{ B_j, \; j=1,\ldots,
2^N- 1 \}$ be a set of homogeneous, linearly independent monomials in
$\wedge(N)$, whose degree is strictly less than $N$. It is easy to see
that the following family $\F_S$ of mutually local formal
distributions is linearly independent over $\Cset[\d_z]$ and that all
their coefficients form a $\Cset$-basis of $S((1,N))'$:
\begin{equation}\lbb{bassn}
A_i(z)= \sum_{n \in \Zset}(x^n A_i) z^{-n-1}, \quad
B_j(z)= (l-N)B_j^0(z) +\d_z \sum_{i=1}^N(B_j\xi_i)^i(z).
\end{equation}
Hence $(S((1,N))', \F_S)$ is a formal distribution Lie superalgebra,
and the corresponding Lie conformal superalgebra $\ov{\F_S}$ has rank
$N2^N$ over $\Cset[\d_z]$.

Let us describe this Lie conformal superalgebra more explicitly.  For
an element $D=\sum_{i=1}^N P_i(\d,\xi)\d_i + f(\d,\xi) \in W_N$, we
define the corresponding notion of divergence:
\begin{equation*}
div\; D=\sum_{i=1}^N(-1)^{p(P_i)}\d_i P_i -\d f \in \Cset[\d]\tt
\wedge(N).
\end{equation*}
The following identity holds in $\Cset[\d]\tt
\wedge(N)$, where $D_1,D_2 \in W_N$ (cf. \ref{lactwn}):
\begin{equation}\lbb{frmdiv}
div\;[{D_1}_{\la}D_2]={D_1}_{\la}(div\;
D_2)-(-1)^{p(D_1)p(D_2)}{D_2}_{-\la-\d} (div\; D_1).
\end{equation}
Therefore,
\begin{equation*}
S_N=\{ D\in W_N \st div\; D=0 \}
\end{equation*}
is a subalgebra of $W_N$.  The Lie conformal superalgebra $S_N$ is
simple for $N\geq 2$; one can check this by using the same argument as
for $W_N$. Also, it is a free $\Cset[\d]$- module of rank
$N2^N$. Furthermore, $(S((1,N))', \F_S)$, where $\F_S$ is defined by
(\ref{bassn}), is the maximal formal distribution Lie superalgebra
associated to $S_N$. This follows from Proposition \ref{companna}. The
above discussion implies that
\begin{equation*}
\A(S_N) = \ov{S}(1,N)' \quad \text{ and} \quad
\A(S_N)^e= \Cset ad(\d_0) \ltimes \ov{S}(1,N)'.
\end{equation*}
\end{example}

\begin{example}\lbb{s2except2}
For any $a \in \Cset$, we set
\begin{equation*}
S((1,N,a))= \{ D \in W((1,N)) \st div \;e^{ax}D=0 \}.
\end{equation*}
This is a subalgebra of $W((1,N))$, which is spanned by the
coefficients of the following family $\F_{S,a}$ of mutually local
formal distributions
\begin{equation}\lbb{bassna}
A_i(z), \quad B_j(z)= (l-N)B_j^0(z) +(\d_z-a) \sum_{i=1}^N(B_j\xi_i)^i(z),
\quad \text{cf. (\ref{bassn})}.
\end{equation}

The associated Lie conformal superalgebra is constructed explicitly as
follows.  Let $D=\sum_{i=1}^N P_i(\d,\xi)\d_i + f(\d,\xi)$ be an
element of $W_N$.  We define the deformed divergence to be
\begin{equation*}
div_a D= div D + a f.
\end{equation*}
It still satisfies formula
(\ref{frmdiv}), hence
\begin{equation*}
S_{N,a}= \{ D \in W_N \st div_a D = 0 \}
\end{equation*}
is a subalgebra of $W_N$, which is simple for $N \geq 2$ and has rank
$N2^N$.

As for the annihilation algebra, for $a\neq 0$ the automorphism of
$\ov{\wedge}(1,N)$ sending $x$ to $\frac{1}{a} (e^{ax}-1)$ and leaving
the $\xi_i$'s invariant induces an automorphism in the space of the
differential forms, which transforms the standard volume form $d x
\wedge v $ into $e^{ax} dx \wedge v$. Hence
\begin{equation*}
\A(S_{N,a})\simeq \ov{S}(1,N)'.
\end{equation*}
Moreover, using Lemmas \ref{dersnnbig} and \ref{ders2}, one can see
that the induced automorphism in $Der(\ov{S}(1,N)')/ad(\ov{S}(1,N)')$
sends $ad(\d_0)$ to $ad(\d_0 -a
\sum_{i=1}^N\xi_i\d_i)$. Consequently,
\begin{equation*}
\A(S_{N,a})^e=\Cset  ad(\d_0 -a
\sum_{i=1}^N \xi_i \d_i)
\ltimes
\ov{S}(1,N)'.
\end{equation*}
\end{example}



\begin{example}\lbb{s2except1}
Let $N \in \Zset_+$ be an even integer. We set
\begin{equation*}
\wti{S}((1,N))=\{ D\in W((1,N)) \st div( (1+ \xi_1 \ldots \xi_N)D )=0 \}.
\end{equation*}
This is a subalgebra of $W((1,N))$, which is spanned by the
coefficients of the following family $\F_{\wti{S}}$ of mutually local
formal distributions
\begin{equation}\lbb{bassnti}
\wti{A}_i(z)= (1- \xi_1\ldots \xi_N)A_i(z), \quad
B_j(z), \quad \text{cf. (\ref{bassn})}.
\end{equation}
The associated Lie conformal superalgebra $\wti{S}_N$ is constructed
explicitly as follows:
\begin{equation*}
\wti{S}_N=\{D \in W_N \st div((1+\xi_1 \ldots \xi_N)D)=0 \}
           (=(1-\xi_1 \ldots \xi_N)S_N).
\end{equation*}
The Lie conformal superalgebra $\wti{S}_N$ is simple for $N \geq 2$
and has rank $N2^N$.

As for the annihilation algebra, we remark that the automorphism of
$\ov{\wedge}(1,N)$ sending $x$ to $(1+ \xi_1 \ldots \xi_N)x$ and
leaving the $\xi_i's$ unchanged induces an automorphism in the space
of differential forms which transforms the standard volume form
$dx\wedge v$ into $(1+ \xi_1 \ldots \xi_N)dx\wedge v$. Hence,
\begin{equation*}
\A(\wti{S}_N)=\ov{S}(1,N)'.
\end{equation*}
Moreover, $ad(\d_0)$ is sent to $ad(\d_0 - \xi_1
\ldots \xi_N \d_0)$, so that
\begin{equation*}
\A(\wti{S}_N)^e=\Cset  ad(\d_0 - \xi_1 \ldots
\xi_N \d_0)  \ltimes \ov{S}(1,N)'.
\end{equation*}
\end{example}


\begin{example}\lbb{kser}
Given the differential form
\begin{equation*}
 \omega=dx -\sum_{i=1}^N \xi_i d\xi_i,
\end{equation*}
we define the following subalgebra of $W((1,N))$
(resp. $\ov{W}(1,N)$):
\begin{align*}
K((1,N))\;(\text{resp.}\;\ov{K}(1,N)) =&\{ D \in
W((1,N))(\text{resp.}\;
\ov{K}(1,N)) \st \\
&D\omega=P\omega \;\; \exists P \in \wedge((1,N))(\text{resp.} \;
\ov{\wedge}(1,N)) \},
\end{align*}
see \cite{K5}. It consists of linear operators of the form $(f \in
\wedge((1,N))$ (resp. $\ov{\wedge}(1,N)$))
\begin{equation*}
D^f=f\d_0 +\frac{1}{2}(-1)^{p(f)}\sum_{i=1}^N(\xi_i \d_0 +\d_i)(f)
(\xi_i\d_0 +\d_i).
\end{equation*}
The Lie superalgebra $\ov{K}(1,N)$ is linearly compact and simple for
all $N\in \Zset_+$ \cite{K5}.  The space $\wedge((1,N))$
(resp. $\ov{\wedge}(1,N)$) can be identified with the Lie superalgebra
$K((1,N))$ (resp. $\ov{K}(1,N)$) via the map $f \to D^f$, in which
case the bracket becomes, for $f,g \in \wedge((1,N))$
(resp. $\ov{\wedge}(1,N)$):
\begin{equation*}
[f,g]=(f- \frac{1}{2}\sum_{i=1}^N \xi_i \d_i f ) \d_0 g -
\d_0 f(g -\frac{1}{2}\sum_{i=1}^N \xi_i \d_i g)
+ (-1)^{p(f)}\frac{1}{2}
\sum_{i=1}^N \d_i f \d_i g.
\end{equation*}

For $A\in \wedge(N)$, we define the $\wedge((1,N))$-valued formal
distribution
\begin{equation*}
A(z)=\sum_{n \in \Zset}(x^n  A)\; z^{-n-1}.
\end{equation*}
If $A=\xi_{i_1} \ldots \xi_{i_r}, \;\; B=\xi_{j_1} \dots \xi_{j_s}$ we
have
\begin{align*}
[A(z),B(w)]=&\left((\frac{r}{2}-1)\d_wAB(w) + (-1)^r
\frac{1}{2}\sum_{i=1}^N (\d_iA\d_iB)(w)\right)\de(z-w) \\
+&\left(\frac{r+s}{2}-2\right)AB(w)\d_w\de(z-w),
\end{align*}
hence the formal distributions in $\F_K=\{ A(z) \}_{A \in \wedge(N)}$
are mutually local and $(K((1,N)), \F_K)$ is a formal distribution Lie
superalgebra. The associated Lie conformal superalgebra is
\begin{equation*}
K_N=\Cset[\d] \tt \wedge(N)
\end{equation*}
with $\la$-bracket
\begin{equation*}
[A_{\la}B]=\left((\frac{r}{2}-1)\d (AB) + (-1)^r
\frac{1}{2}\sum_{i=1}^N
\d_iA\d_iB\right) +\la\left(\frac{r+s}{2}-2\right)AB.
\end{equation*}
Using the same argument as for $W_N$, one can see that the Lie
conformal superalgebra $K_N$ is simple for all $N \in \Zset_+,
\; N \neq 4$ and is a free $\Cset[\d]$-module of rank $2^N$. By
Proposition \ref{companna}, we have
\begin{equation*}
\A(K_N)=\ov{K}(1,N) \quad \text{and } \quad
\A(K_N)^e= \Cset ad(\d_0) \ltimes \ov{K}(1,N).
\end{equation*}

\end{example}


\begin{example}\lbb{k'4ex}
The Lie superalgebra $K((1,4))$ is not simple:
\begin{displaymath}
K((1,4))=K((1,4))' \oplus (\Cset x^{-1}
\xi_1
\ldots \xi_4),
\end{displaymath}
but $K((1,4))'=[K((1,4)),K((1,4))]$ is simple.  Also, $K((1,4))'$ is a
formal distribution Lie superalgebra, spanned by the coefficients of
the family of mutually local formal distributions
\begin{equation*}
\F_{K'}=\{ A(z), \; A \;\; \text{a monomial in} \wedge(4), A\neq \xi_1
\ldots \xi_4;
    \; \d_z \xi_1 \ldots \xi_4(z) \}.
\end{equation*}

Its associated Lie conformal superalgebra is $K'_4$, the derived
subalgebra of $K_4=K'_4 \oplus \Cset \xi_1 \ldots \xi_4$. By formula
(\ref{brinannia}), $(\d \xi_1 \ldots \xi_4)_0$ is a central element of
the formal distribution Lie superalgebra $(Lie \; K'_4,
\F(K'_4))$. Recall that we have a surjective homomorphism $Lie \; K'_4
\to K((1,4))'$.  It follows from Proposition
\ref{companna} that $Lie \; K'_4$ is a central extension of
$K((1,4))'$ by a $1$-dimensional center. We denote it by
$CK((1,4))'$. The corresponding cocycle is given by \cite{KL}, formula
$(4.22)$, for $d=0$. It follows that the annihilation algebra of
$K'_4$ is a central extension $C\ov{K}(1,4)$ of $\ov{K}(1,4)$ obtained
by restricting the above central extension of $K((1,4))'$ to the
subalgebra $K(1,4)$ and then going to the completion
$\ov{K}(1,4)$. The non-zero entries of the corresponding cocycle are
given by:
\begin{equation}\lbb{cocyk'4}
\psi(1,\xi_1 \ldots \xi_4)=1,  \;\; \psi(\xi_i,\d_i
\xi_1 \ldots \xi_4)
=\frac{1}{2},\;\; i=1,2,3,4.
\end{equation}
Also, $\A(K'_4)^e=\Cset ad(\d_0) \ltimes C\ov{K}(1,4)$, where $\d_0$
acts trivially on the center.

\end{example}

\begin{example}\lbb{ck6exa}
The formal distribution Lie superalgebra $(K((1,6)), \F_K)$ has a
simple subalgebra, denoted by $(CK((1,6)), \F_{CK})$.  The associated
Lie conformal superalgebra is $CK_6$. It is a simple rank 32
subalgebra of $K_6$, whose even part is $W_0 \ltimes Cur\;
\so_6$ and whose odd part is spanned by six primary fields of
conformal weight 3/2 and ten primary fields of conformal weight
1/2. For the explicit form of the commutation relations, as well
as for more detailed information on $CK_6$, see \cite{CK2}.

The annihilation algebra of $CK_6$ is the exceptional simple linearly
compact Lie superalgebra $\ov{E}(1,6)$ (see \cite{CK2, K5}), which is
a subalgebra of $\ov{K}(1,6)$. The extended annihilation algebra is
$\A(CK_6)^e=\Cset \d_0
\ltimes \ov{E}(1,6)$.
\end{example}

\begin{remark}
The Virasoro conformal algebra is the only non-abelian rank 1 Lie
conformal algebra. One has $W_0=\Cset[\d]L$, where $[L_{\la}L]=(\d
+2\la)L$.  An even element $L$ of a Lie conformal superalgebra
satisfying this $\la$-bracket is called a Virasoro element.
\end{remark}

\begin{remark}
The following are the most important Lie conformal superalgebras.
$W_0 \simeq K_0$ is the Virasoro conformal algebra, $K_1$ is the
Neveu-Schwarz Lie conformal superalgebra, $K_2 \simeq W_1$ is the
$N=2$ Lie conformal superalgebra, $K_3$ is the $N=3$ Lie conformal
superalgebra, $S_2$ is the $N=4$ Lie conformal superalgebra, and
$K'_4$ is the big $N=4$ Lie conformal superalgebra. The corresponding
formal distribution Lie superalgebras (or, rather, their central
extensions) are well known and play an important role in physics.
\end{remark}

\section{Central Extensions}\lbb{scenext}

\subsection{Central extensions of Lie superalgebras}\lbb{subcohls}

We shall be dealing with the cohomology $H^2(\g)$ of a Lie
superalgebra $\g$ with coefficients in the trivial $\g$-module
$\Cset$, which parameterizes the central extensions of $\g$. The
following lemma is well known (cf. \cite{K1}, Exercise 7.6).

\begin{lemma}\lbb{mph2h1}
Let $\g$ be a finite-dimensional Lie superalgebra, which has an
invariant, supersymmetric non-degenerate bilinear form (even or
odd). Then the $2$-cocycles on $\g$ with coefficients in $\Cset$ are
in $1-1$ correspondence with the derivations of $\g$ that are
skew-symmetric with respect to the form, the cocycle being trivial iff
the corresponding derivation is inner. This correspondence is given by
$\al_D(x,y)=(Dx,y)$, $D \in Der(\g)$, $x,y
\in \g$.
\end{lemma}

%
\begin{proof}
If $\al$ is a $2$-cocycle on $\g$, it can be written in the form
$\al_D(x,y)=(Dx,y)$, where $D$ is an endomorphism of the space $\g$.
One can easily check that the skew-symmetry of $\al$ is equivalent to
the skew-symmetry of $D$, that the cocycle equation is equivalent to
$D$ being a derivation and that $\al$ is a trivial cocycle iff $D$ is
an inner derivation.
\end{proof}

In the remaining part of this Subsection, we compute the central
extensions of all simple finite-dimensional Lie superalgebras $\ss$
and their $\Cset[[t]]$-current algebras $\ss[[t]]$.

The main result in \cite{K4} is the following theorem.

\begin{theorem}
A simple finite-dimensional Lie superalgebra is isomorphic either to
one of the simple Lie algebras or to one of the Lie superalgebras
$A(m,n)$ ($0\leq m < n$),
$A(n,n)$ ($n>0$),
$B(m,n)$ ($m\geq 0$, $n>0$),
$C(n)$ ($n\geq 3$),
$D(m,n)$ ($m\geq 2$, $n>0$),
$D(2,1,\alpha)$, $F(4)$, $G(3)$,
$P(n)$ ($n\geq 2$),
$Q(n)$ ($n\geq 2$),
$W(n)$ ($n\geq 3$),
$S(n)$ ($n\geq 4$),
$\wti{S}(n)$ ($n\geq 4$, n even),
or $H(n)$ ($n\geq 5$). \hfill $\Box$
\end{theorem}

\begin{proposition}\lbb{invbilfor}
A simple finite-dimensional Lie superalgebra $\ss$ has an even
(resp. odd) supersymmetric, invariant bilinear form iff $\ss$ is
isomorphic to $A(m,n)$, $B(m,n)$, $C(n)$, $D(m,n)$, $D(2,1,\alpha)$,
$G(3)$, $F(4)$, $H(n)$ with n even (resp. $\ss \simeq H(n)$ with $n$
odd or $Q(n)$).  Such a form is unique up to a constant factor.
\end{proposition}

\begin{proof}
The Lie superalgebras $A(m,n)$, $B(m,n)$, $C(n)$, $D(m,n)$,
$D(2,1,\alpha)$, $G(3)$ and $F(4)$ are contragredient with a
symmetrizable Cartan matrix, hence they have an even invariant form
(\cite{K4}).

The obvious pairing between the even and the odd part of $Q(n)$ is an
odd invariant form.

Define on $\wedge(n)$ the Poisson bracket
\begin{equation*}
\{f,g\}=\sum_{i=1}^n\frac{\d f}{\d \xi_i}\frac{\d g}{\d \xi_i}.
\end{equation*}
This Lie superalgebra has a bilinear form $(f,g)=\int fg \;
d\xi_1\ldots d\xi_n$ (where $\int$ denotes the Berezin integral,
cf. \cite{Be}), which is invariant on the derived subalgebra
$\wedge(n)'$ i.e. the span of all monomials except the top one. The
kernel of the restriction of the form to $\wedge(n)'$ is
$\Cset$. Since $H(n)\simeq\wedge(n)'/\Cset$, the proposition is proved
in this case as well.

In order to show that in the remaining cases there is no invariant
form on $\ss$, we take a maximal reductive Lie subalgebra $\mathfrak
r$ of $\ss$ and show that the $\mathfrak r$-modules $\ss$ and
$\ss^{\star}$ are not isomorphic.
\end{proof}

\begin{proposition}\lbb{h2fdls}
A complete list of non-trivial $H^2(\ss)$ for all simple
finite dimensional Lie superalgebras $\ss$ is as follows:
\begin{align*}
&H^2(A(1,1))=H_{\text{even}}^2(A(1,1))= \Cset^3; \\
&H^2(A(n,n))=H_{\text{even}}^2(A(n,n))= \Cset, \quad n>1; \\
&H^2(P(3))=H_{\text{even}}^2(P(3))= \Cset; \\
&H^2(Q(n))=H_{\text{even}}^2(Q(n))=\Cset,
\quad n\geq 2; \\
&H^2(H(n))=H_{\text{even}}^2(H(n))=\Cset,\;\; n\geq 5.
\end{align*}
\end{proposition}

\begin{proof}
In the cases when $\ss$ carries an invariant, supersymmetric,
non-degenerate bilinear form, we apply Lemma \ref{mph2h1} and the
known description of derivations (see \cite{K4} Proposition
5.1.2). Notice that in all cases except $H(n)$ all derivations are
skew-symmetric, whereas for $H(n)$, $n\geq 5$, only one of the two
outer derivations, namely $D=\sum_i(\d_i \xi_1\ldots\xi_n )\d_i$, has
this property.  In the remaining cases, i.e. $\ss$=$P(n)$, $S(n)$,
$\wti{S}(n)$ and $W(n)$, we apply Lemma 2.1 of \cite{KL} to compute
$H^2(\ss)$.
\end{proof}

\begin{remark}
The central extensions corresponding to the non-trivial cocycles
listed above are as follows:
\begin{enumerate}
\item $A(n,n)$: the canonical homomorphism $\sl(n+1,n+1) \to A(n,n)$;
\item $Q(n)$: the canonical homomorphism $\wti{Q}(n) \to Q(n)$;
\item if $\ss=\oplus_{j \geq -1}\ss_j$ (resp. $\ss=\oplus_{j \leq 1}\ss_j$)
is a consistent $\Zset$-gradation of $\ss$ such that the
$\ss_0$-module $\ss_{-1}$ carries a non-zero symmetric invariant
bilinear form $(.,.)$, then we have the following 2-cocycle on $\ss$:
$\al(x,y)=(x,y)$, if $x,y
\in \ss_{-1}$ (resp. $x,y \in \ss_{+1}$), and $\al(x,y)=0$ if $x\in
\ss_i$, $y\in \ss_j$, and $i$ or $j$ is not $-1$ (resp. $i$ or $j$ is
not $+1$); the central extensions of $H(n)$, $P(3)$ and the remaining
two central extensions of $A(1,1)$ are obtained in this way.
\end{enumerate}
\end{remark}

\begin{proposition}\lbb{cohcurr}
Let $\ss$ be a simple finite-dimensional Lie superalgebra. Then any
irreducible central extension of $\ss[[t]]$ is isomorphic to
$\what{\ss}[[t]]$, where $\what{\ss}$ is an irreducible central
extension of $\ss$.
\end{proposition}
\begin{proof}
Note that all K\"ahler differentials of $\Cset[[t]]$ are exact. Hence
one can use the same argument as in \cite{Sa}.
\end{proof}

\subsection{Central extensions of Lie conformal superalgebras}\lbb{subcohlcs}

Basic and reduced cohomology of Lie conformal algebras are defined in
\cite{BKV}. They are denoted respectively by $\wti{H^{\star}}$ and
$H^{\star}$. The same definitions (with appropriate signs) apply
to the case of Lie conformal superalgebras.

A central extension of a Lie conformal superalgebra $R$ by a Lie
conformal superalgebra $K$ is a $\Cset[\d]$-split (i.e. $S=K\oplus R$
as $\Cset[\d]$-modules) short exact sequence of Lie conformal
superalgebras
\begin{equation*}
0 \to K \to  S  \to R \to 0,
\end{equation*}
in which $K$ is central in $S$ and the action of $\d$ is trivial on
$K$. Equivalence classes of $1$-dimensional central extensions are
parameterized by elements of the second reduced cohomology group
$H^2(R)$ (cf. \cite{BKV}).

\begin{lemma}\lbb{deccoh}
Let $R$ be a Lie conformal superalgebra. Suppose in $R$ there is an
element $L$ such that $L_{(0)}a=\d a$ for any $a\in R$. Then
\begin{equation*}
H^n(R)=H^n(\A(R))\oplus H^{n+1}(\A(R)), \quad  n\geq 0.
\end{equation*}
\end{lemma}


\begin{proof}
This follows from \cite{BDK}, Proposition 15.6.
\end{proof}


\begin{lemma}\lbb{peimh1red0}
Let $R$ be a Lie conformal superalgebra such that $R=R'$. Then
$H^1(R)=0$.
\end{lemma}


\begin{proof}
We denote by $\wti{d}^n$ the $n$-th differential of the basic complex
and by $d^n$ the corresponding differential of the reduced complex.
$\Cset$ is a trivial $\Cset[\d]$-module, hence $\wti{d}^0=d^0=0$, and
$H^1(R)=ker d^1$.

We remark that $\gamma \in \d \wti{C}^1(R)$ if, and only if, for any
$a \in R$, $\gamma_{\la}(a)$ has no constant term as polynomial in
$\la$.  Indeed, if $\gamma=\d \omega$, $\gamma_{\la}(a)=(\d
\omega)_{\la}(a)=\la \omega_{\la}(a)$.  Also, if $\gamma_{\la}(a)=\la
P_a(\la)$ for any $a \in R$, we can define $\omega \in \wti{C}^1(R)$
such that $\omega_{\la}(a)=P_a(\la)$ and clearly $\gamma = \d \omega$.

Recall that
\begin{equation*}
d^1 : \;\; \frac{\wti{C}^1(R)}{\d \wti{C}^1(R)} \to
\frac{\wti{C}^2(R)}{\d \wti{C}^2(R)}.
\end{equation*}
Suppose $[\gamma]=\gamma + \d \wti{C}^1(R) \in ker d^1$.  Then
$d^1([\gamma])=0$, i.e. $\wti{d}^1 \gamma \in \d \wti{C}^2(R)$.  It
follows that, for any $a_1,a_2 \in R$,
\begin{equation*}
(\wti{d}^1 \gamma)_{\la_1,\la_2}(a_1,a_2)= (\d
\beta)_{\la_1,\la_2}(a_1,a_2)= (\la_1 +\la_2)
\beta_{\la_1,\la_2}(a_1,a_2).
\end{equation*}
On the other hand, $(\wti{d}^1 \gamma)_{\la_1,\la_2}(a_1,a_2)=-
\gamma_{\la_1 +\la_2}([{a_1}_{\la_1}a_2])$.
Therefore, the polynomial $\gamma_{\la}(a)$ has no constant term for
any $a \in R'$. But $R=R'$, so this actually holds for any $a \in R$.
It follows that $\gamma \in \d \wti{C}^1(R)$, i.e. $[\gamma]=0$.
\end{proof}

\begin{lemma}\lbb{decseccoh}
Let $R$ be a Lie conformal superalgebra, such that $R=R'$. Suppose $R$
has an element $L$ such that $L_{(0)}a=\d a$ for any $a\in R$. Then
$H^2(\A(R))= 0$. In particular, this holds for the annihilation
algebras of the simple Lie conformal superalgebras $W_N$ ($N\geq 0$),
$S_{N,a}$ ($N\geq 2$), $\wti{S}_N$ ($N$ even, $N\geq 2$), $K_N$
($N\geq 0$, $N\neq 4$), $K'_4$, $CK_6$.
\end{lemma}

\begin{proof}
This follows from Lemmas \ref{deccoh} and \ref{peimh1red0}, since
$0=H^1(R)=H^1(\A(R))\oplus H^2(\A(R))$.
\end{proof}

\begin{lemma}\lbb{invsecredcoh}
Let $R$ be a finite Lie conformal superalgebra, which is free as a
$\Cset[\d]$-module. Let $\ss$ be a reductive Lie algebra, and let $\ss
\to R$ be an injective homomorphism with respect to the $0$-th product
on $R$. Assume that $R=\oplus_{i\in I}\g_i$ is a decomposition of $R$
into a direct sum of finite-dimensional, irreducible $\ss$-modules
with respect to the $0$-th product. Then every reduced $2$-cocycle
$\psi$ on $R$ is equivalent to a cocycle $\al_{\la}$ such that
\begin{enumerate}
\item $\al_{\la}(\g_i,\g_j)=0$ if $\g_i,\;\g_j$ are not
contragredient $\ss$-modules;
\item  $\al_{\la}(x,y)=P_{ij}\langle x,y \rangle$ for all $x\in \g_i, \;
y \in \g_j$, and some $P_{ij} \in \Cset[\la]$ if $\g_i,\;\g_j$ are
contragredient $\ss$-modules $(\langle,\rangle$ denotes the pairing
between them$)$.
\end{enumerate}
\end{lemma}

\begin{proof}
The Lie algebra $\ss$ acts completely reducibly on the space of
reduced $2$-cocycles, hence there exists an $\ss$-invariant subspace
$V$ complementary to the space of trivial $2$-cocycles. Since $R$ acts
trivially on $H^2(R)$, we conclude that $\ss$ acts trivially on $V$.
\end{proof}

In the remaining part of this Section, we compute the central
extensions of all simple Lie conformal superalgebras listed in Section
\ref{subsomeex}.
The proofs are all based on Lemma \ref{invsecredcoh}. We give all the
details of the computations only in the most involved case
(cf. Proposition \ref{h2lieconsn}) and omit them in all other cases.
We give only the non-zero entries of the non-trivial cocycles.

\begin{proposition}\lbb{h2concurr}
Let $\ss$ be a simple finite-dimensional Lie superalgebra. Then all
central extensions of $Cur\; \ss$ are given by the following
$2$-cocycles
\begin{equation*}
\al_{\la}(a,b)=\al_0(a,b) +(a,b)\la, \;\;\; a,b \in \ss,
\end{equation*}
where $\al_0(a,b)$ is a 2-cocycle on $\ss$ and $(.,.)$ is a
supersymmetric, invariant bilinear form on $\ss$ (cf. Propositions
\ref{invbilfor} and \ref{h2fdls}). Two such cocycles
are equivalent iff the 2-cocycles $\al_0$ are equivalent.
\end{proposition}

\begin{proof}
See \cite{K2}, Section 2.7.
\end{proof}

In what follows we shall often denote an element $1\tt (\sum_{i=1}^N
f_i \d_i +f) \in W_N= \Cset[\d]\tt(W(N) \oplus \wedge(N))$ by $\sum_i
(f_i)^i +f$.


\begin{proposition}
The Lie conformal superalgebras $W_0$, $W_1$ and $W_2$ have a unique,
up to isomorphism, central extension. The corresponding 2-cocycles are
as follows:
\begin{alignat*}{4}
&W_0:\;\;&&\alpha_3(1,1)=\frac{1}{2};&& \\
&W_1:\;\;&&\alpha_1((\xi_1)^1,(\xi_1)^1)=\frac{1}{3},\;\;&
&\alpha_2((\xi_1)^1,1)=\frac{1}{3},\;\;&&
\alpha_2((1)^1,\xi_1)=-\frac{1}{3}; \\
&W_2:\;\;&&\alpha_1((\xi_1)^2,(\xi_2)^1)=\frac{1}{6},\;\;&
&\alpha_1((\xi_1)^1,(\xi_2)^2)=-\frac{1}{6},\;\;
&&\alpha_1((\xi_1\xi_2)^2,(1)^1)=-\frac{1}{6}, \\
&&&\alpha_1((\xi_1\xi_2)^1,(1)^2)=\frac{1}{6}.
\end{alignat*}
\end{proposition}
\begin{proposition}
The Lie conformal superalgebra $W_N$ has no non-trivial central
extensions if $N\geq 3$.
\end{proposition}
\vspace{1mm}
Using \cite{AF} and Lemma \ref{deccoh} one actually computes the whole
basic and reduced cohomology of $W_N$.
\vspace{1mm}
\begin{proposition}
The Lie conformal superalgebra $S_{2,a}$ has a unique, up to
isomorphism, non-trivial central extension. The corresponding
2-cocycle is as follows:
\begin{alignat*}{2}
&\al_3(L_a,L_a)=\frac{1}{2}, \qquad \text{where} \;\; &&L_a =-1 +
\frac{1}{2} (\d -a)((\xi_1)^1 +(\xi_2)^2), \\ &\al_2((1)^1,
\xi_1-(\d-a)(\xi_1\xi_2)^2 )=-\frac{1}{3}, &&\al_1((1)^1,
\xi_1-(\d-a)(\xi_1\xi_2)^2 )=\frac{a}{6}, \\ &\al_0((1)^1,
\xi_1-(\d-a)(\xi_1\xi_2)^2 )=-\frac{a^2}{24},
\;\;  &&\al_2((1)^2, \xi_2 + (\d -a)(\xi_1\xi_2)^1)=-\frac{1}{3},  \\
&\al_1((1)^2, \xi_2 + (\d-a)(\xi_1\xi_2)^1)=\frac{a}{6},
&&\al_0((1)^2,\xi_2 + (\d-a) (\xi_1\xi_2)^1)=-\frac{a^2}{24}, \\
&\al_1((\xi_1)^2,(\xi_2)^1)=\frac{1}{6},
&&\al_1((\xi_1)^1-(\xi_2)^2, (\xi_1)^1-(\xi_2)^2)=\frac{1}{3}.
\end{alignat*}
\end{proposition}

\begin{proposition}
The Lie conformal superalgebra $\wti{S}_2$ has a unique, up to
isomorphism, non-trivial central extension. The corresponding
2-cocycle is as follows:
\begin{alignat*}{2}
&\alpha_3(\wti{L}, \wti{L})=\frac{1}{2}, \qquad
\text{where} \;\;&&\wti{L}=-(1-\xi_1\xi_2) +
\frac{1}{2}\d ((\xi_1)^1 +(\xi_2)^2), \\
&\alpha_1((1-\xi_1\xi_2)^1,(1-\xi_1\xi_2)^2)=-\frac{1}{3}, \quad
&&\alpha_2((1-\xi_1\xi_2)^2,\xi_2 + \d(\xi_1\xi_2)^1)=-\frac{1}{3}, \\
&\alpha_2((1-\xi_1\xi_2)^1,\xi_1 -
\d(\xi_1\xi_2)^2)=-\frac{1}{3},  \quad
&&\alpha_1((\xi_1)^2,
(\xi_2)^1)=\frac{1}{6}, \\
&\alpha_1((\xi_1)^1-(\xi_2)^2,(\xi_1)^1-(\xi_2)^2)= \frac{1}{3}.
\end{alignat*}
\end{proposition}

\begin{proposition}\lbb{h2lieconsn}
The Lie conformal superalgebras $S_{N,a}$ $(a\in \Cset)$ and
$\wti{S}_N$ ($N$ even) have no non-trivial central extensions if
$N>2$.
\end{proposition}

\begin{proof}
We remark that $S_N \simeq S_{N,a} \simeq \wti{S}_N \simeq W(N)$ as
$\sl_N$-modules.  By Lemma \ref{invsecredcoh}, we need to compute the
cocycles corresponding to the $\sl_N$-invariants of $W(N) \tt W(N)$,
which occur in the following cases $(1\leq k \leq N-1)$:
\begin{equation*}
R(\pi_1 +\pi_{N-1}) \tt R(\pi_1 +\pi_{N-1}), \; R(0)\tt R(0), \;
R(\pi_{N-1}) \tt R(\pi_1), \; R(\pi_k) \tt R(\pi_{N-k}).
\end{equation*}
Let $L_a=-1 + \frac{1}{N}(\d-a)H$
(resp. $\wti{L}=-(1-\xi_1\ldots\xi_N) + \frac{1}{N} \d H$) be a
Virasoro element for $S_{N,a}$ (resp. $\wti{S}_N$), where
$H=\sum_{i=1}^N \xi_i \d_i$ denotes the Euler operator. Let $\d_N$
(resp. $(1-\xi_1\ldots\xi_N)\d_N$) be the highest weight vector of
$R(\pi_{N-1}) \subseteq W(N)_{-1}$ in $S_{N,a}$ (resp. $\wti{S}_N$).
Let $v^a_k$ (resp. $\wti{v_k}$) be the highest weight vector of
$R(\pi_k)
\subseteq W(N)_{k}$ in $S_{N,a}$
(resp. $\wti{S}_N$), $1\leq k \leq N-1$. Then
\begin{equation*}
v^a_k= (k-N)\xi_1\ldots \xi_k +(\d-a)\xi_1\ldots \xi_kH,  \quad
\wti{v_k}=(k-N)\xi_1\ldots \xi_k +\d\xi_1\ldots \xi_kH.
\end{equation*}
Let $w^a_k$ (resp. $\wti{w_k}$) be the lowest weight vector of
$R(\pi_{N-k}) =R(\pi_k)^{\star}$, which we view as lowest weight
module $R(-\pi_k)^{\text{low}}\subseteq W(N)_{N-k}$ in $S_{N,a}$
(resp. $\wti{S}_N$), $1\leq k \leq N-1$. Then
\begin{equation*}
w^a_k= -k\xi_{k+1}\ldots \xi_N +(\d-a)\xi_{k+1}\ldots \xi_NH, \quad
\wti{w_k}=-k\xi_{k+1}\ldots \xi_N + \d \xi_{k+1}\ldots \xi_NH.
\end{equation*}
The action of $L_a$ and $\wti{L}$ is given by the following formulas
$(1\leq k \leq N-1)$:
\begin{alignat*}{2}
&[{L_a}_{\la}\d_N]=\frac{1}{N}(N\d +(N+1)\la +a )\d_N,  \;&&\\
&[\wti{L}_{\la}((1-\xi_1\ldots \xi_N)\d_N)]= \frac{1}{N}(N\d +(N+1)\la)
(1&&-\xi_1\ldots\xi_N)\d_N -v^0_{N-1}, \\
&[{L_a}_{\la}v^a_k]=\frac{1}{N}(N\d +(2N-k)\la -ak )v^a_k,  &
&[\wti{L}_{\la}\wti{v_k}]=\frac{1}{N}(N\d +(2N-k)\la)\wti{v_k}, \\
&[{L_a}_{\la}w^a_k]=\frac{1}{N}(N\d +(N+k)\la -a(N-k) )w^a_k, &
&[\wti{L}_{\la}\wti{w_k}]=\frac{1}{N}(N\d +(N+k)\la)\wti{w_k}.
\end{alignat*}

We remark that $R(\pi_1 + \pi_{N-1}) \subseteq Cur \;S(N)$ for
$S_{N,a}$ and $R(\pi_1 + \pi_{N-1}) \subseteq Cur \;\wti{S}(N)$ for
$\wti{S}_N$. By Proposition \ref{h2concurr}, the corresponding cocycle
is trivial.

In all other cases, by Lemma \ref{invsecredcoh} we need to show that
the cocycle pairing the highest weight vector and the lowest weight
vector is trivial.  The cocycle equation for the triples $(v^a_k,
\xi_k\d_k-\xi_{k+1}\d_{k+1},w^a_k)$, $(L_a,v^a_k,w^a_k)$ and
$(\wti{v_k}, \xi_k\d_k-\xi_{k+1}\d_{k+1},\wti{w_k})$,
$(\wti{L},\wti{v_k},\wti{w_k})$ shows that
$\al_{\la}(v^a_k,w^a_k)=\al_{\la}(\wti{v_k},\wti{w_k})=0$.

Finally, the cocycle equation for the triple $(L_a,\d_N,w^a_1)$ shows
that $\al_{\la}(L_a,L_a)$ and $\al_{\la}(\d_N,w^a_1)$ are trivial.
Similarly, one can see that the cocycle equation for the triple
$(\wti{L},(1-\xi_1\ldots\xi_N)\d_N,\wti{w_1})$ implies that
$\al_{\la}(\wti{L},\wti{L})$ and
$\al_{\la}((1-\xi_1\ldots\xi_N)\d_N,\wti{w_1})$ are also trivial.
Therefore, $H^2(S_{N,a})=H^2(\wti{S}_N)=0$.
\end{proof}

\begin{proposition}
The Lie conformal superalgebras $K_0$, $K_1$, $K_2$ and $K_3$ have a
unique, up to isomorphism, central extension. The corresponding
$2$-cocycles are as follows:
\begin{alignat*}{2}
&   \quad &&\alpha_3(1,1)=\frac{1}{2}, \quad
 \alpha_2(\xi_i,\xi_i)=\frac{1}{6}
\quad (i=1,2), \quad
 \alpha_1(\xi_1\xi_2,\xi_1\xi_2)=-\frac{1}{12},  \\
& \quad  && \alpha_1(\xi_i\xi_j,\xi_i\xi_j)=-\frac{1}{12} \quad (i\neq j),
\quad \alpha_0(\xi_i\xi_j\xi_k,\xi_i\xi_j\xi_k)=-\frac{1}{12} \quad
(i\neq
j\neq k).
\end{alignat*}
\end{proposition}


\begin{proposition}
The Lie conformal superalgebra $K_4'$ has two, up to isomorphism,
central extensions. The corresponding $2$-cocycles are as follows:
\begin{alignat*}{3}
&\alpha_3(1,1)=\frac{1}{2},&&\alpha_2(\xi_i,\xi_i)=\frac{1}{6},&
&\alpha_1(\xi_i\xi_j,\xi_i\xi_j)=-\frac{1}{12}   \quad (i\neq j),  \\
&\alpha_0(\xi_i\xi_j\xi_k,\xi_i\xi_j\xi_k)=-&&
\frac{1}{12}\; (i\neq j\neq k),&&\alpha_1(\d \xi_1\xi_2\xi_3\xi_4,
\d\xi_1\xi_2\xi_3\xi_4)=-\frac{1}{12}; \\
&\be_2(1,\d \xi_1\xi_2\xi_3\xi_4)=2, &
&\be_1(\xi_i,\d_i(\xi_1\xi_2\xi_3\xi_4))=1,  &
\;&\be_1(\xi_i\xi_j,\d_i\d_j(\xi_1\xi_2\xi_3\xi_4))=-1.
\end{alignat*}
\end{proposition}

\begin{proposition}
The Lie conformal superalgebras $K_N$ ($N\geq 5$) and $CK_6$ have no
non-trivial central extensions.
\end{proposition}

\section{Finite Simple Lie Conformal Superalgebras}\lbb{fsliecosu}

\subsection{The annihilation algebra}\lbb{scannia}

In this Section we study the annihilation algebra $\A(R)$ of a finite
simple Lie conformal superalgebra $R$. We shall use the following two
Propositions, whose proof is the same as in the non-super case, see
\cite{DK}, Lemma 4.3 and Proposition 5.1.

\begin{proposition}\lbb{nocnozr}
Let $R$ be a finite Lie conformal superalgebra. Then any non-central
$T$-invariant ideal $J$ of $\A(R)$ contains a non-zero regular
ideal. \hfill $\Box$
\end{proposition}

\begin{proposition}\lbb{dkrecthm}
Let $R_1$ and $R_2$ be two finite Lie conformal superalgebras, which
are free as $\Cset[\d]$-modules. Let $\varphi: \Cset T_1 \ltimes
\A(R_1) \to \Cset T_2 \ltimes \A(R_2)$ be a homomorphism of
the corresponding extended annihilation algebras such that
$\varphi(T_1)=T_2$ and $\varphi(\A(R_1))=\A(R_2)$. Then there exists a
unique Lie conformal superalgebra homomorphism $\wti{\varphi}: R_1
\to R_2$ that induces $\varphi$, i.e.
$\varphi(a_i)=(\wti{\varphi}(a))_i$ for all $a\in R$, $i \in \Zset_+$.
\hfill $\Box$
\end{proposition}
\begin{lemma}\lbb{annauc}
Let $R$ be a finite simple Lie conformal superalgebra. Then $\A(R)$ is
isomorphic (as a topological Lie superalgebra) to an irreducible
central extension of the Lie superalgebra $\Cset[[t_1, \ldots, t_r]]
\what{\tt} \ss$, where $r =0 \;\text{or}\; 1$ and $\ss$ is a simple
linearly compact Lie superalgebra.
\end{lemma}

\begin{proof}
Recall that $\A(R)$ is a linearly compact Lie superalgebra and that it
is a closed ideal of codimension $1$ in the extended annihilation
algebra $\A(R)^e=\Cset T \ltimes \A(R)$. By Proposition \ref{nocnozr},
due to the simplicity of $R$, $\A(R)$ contains no non-central
$T$-invariant ideals different from $\A(R)$. Let $Z$ be the center of
$\A(R)$. Since the derived algebra of $\A(R)$ is a non-central
$T$-invariant ideal of $\A(R)$ (otherwise $R$ would be nilpotent), we
conclude that $\A(R)=[\A(R),\A(R)]$ and therefore
\begin{equation*}
0 \to Z \to \A(R) \to \A(R)/Z \to 0
\end{equation*}
is an irreducible central extension.  Also, the center of $\A(R)/Z$ is
zero, since otherwise its pre-image in $\A(R)$ would be a proper
non-central $T$-invariant nilpotent ideal.

It follows that $\A(R)/Z$ contains no non-trivial $T$-invariant ideals
and therefore $\A(R)/Z$ is a minimal ideal in $\A(R)^e/Z$. Thus we may
apply Corollary \ref{cgused} to the Cartan-Guillemin theorem to obtain
the isomorphism of linearly compact Lie superalgebras $\A(R)/Z \simeq
(\Cset[[t_1,\ldots,t_r]]\tt
\wedge(m))\what{\tt} \ss$, where $r,m \in \Zset_+$ and $\ss$ is a
simple linearly compact Lie superalgebra.

Next, we show that $m=0$.  Let $\wedge_1(m)$ be the ideal of
$\wedge(m)$ generated by $\xi_1,\ldots,\xi_m$.  We will show that
$I:=(\Cset[[t_1,\ldots,t_r]]\tt \wedge_1(m))\what{\tt} \ss$ is a
$T$-invariant ideal of $\A(R)/Z$, which, of course, will imply that
$m=0$.  Indeed, due to Proposition \ref{detp}, any continuous
derivation $T$ of $(\Cset[[t_1,\ldots,t_r]]\tt \wedge(m))\what{\tt}
\ss$ has the form
\begin{equation}
T=D_0\tt 1 +\sum_i f_i \tt D_i,
\end{equation}
where $D_0$ (resp. $D_i$) is a continuous derivation of
$\Cset[[t_1,\ldots,t_r]]\tt \wedge(m)$ (resp. $\ss$) and $f_i \in
\Cset[[t_1,\ldots,t_r]]\tt \wedge(m)$. Since $T$ is even,  $D_0$ is
an even derivation, hence $T$ maps the ideal $I$ into itself.

It remains to show that $r\leq1$. If $D_0$ leaves the ideal
$(t_1,\ldots,t_r)$ of $\Cset[[t_1,\ldots,t_r]]$ invariant, then
$(t_1,\ldots,t_r)\what{\tt} \ss$ is a non-trivial $T$-invariant ideal
of $\A(R)/Z$, which is impossible.  Therefore there exists a
continuous automorphism of $Der(\Cset[[t_1,\ldots,t_r]])$ which
transforms $D_0$ to $\frac{\d}{\d t_1}$ (see e.g. Proposition
\ref{derconj}). But in this case $T=\frac{\d}{\d t_1} \tt 1 +\sum_i
f_i \tt D_i$, hence the ideal $(t_2,\ldots,t_r)\what{\tt} \ss$ of
$\A(R)/Z$ is $T$-invariant.  This implies that $r \leq 1$.
\end{proof}

\begin{corollary}\lbb{canniauc}
Let R be a finite simple Lie conformal superalgebra and let $L$ be an
even element such that $L_{(0)}a=\d a$ for any $a \in R$. Then $\A(R)$
is the universal central extension of its quotient by the center. In
particular, this holds if $R$ is one of the Lie conformal
superalgebras $W_N$ $(N\geq 0)$, $S_{N,a}$ $(N \geq 2, \; a \in
\Cset)$, $\wti{S}_N$ $(N \geq 2, \; N \text{even})$, $K_N$ $(N \geq0,
\;\; N\neq 4)$, $K'_4$ or $CK_6$.

\end{corollary}
\begin{proof}
As we have just remarked, $\A(R)=[\A(R),\A(R)]$, and by Lemma
\ref{decseccoh}, $H^2(\A(R))=0$.
\end{proof}

By taking the completion  of  (\ref{filanna}), (\ref{qfilanna}),
(\ref{dimqufi}) and (\ref{conoft}) we get:
\begin{equation}\lbb{ficanna}
\A(R) =\ov{\L}_0 \supseteq \ov{\L}_1 \supseteq \ov{\L}_2 \supseteq \ldots,
\end{equation}
\begin{equation}
[\ov{\L}_i,\ov{\L}_j] \subset \ov{\L}_{i+j-d}
\quad \text{for some} \quad d \in \Zset_+,
\end{equation}
\begin{equation}\lbb{dicoquofi}
dim \:\ov{\L}_i/\ov{\L}_{i+1} \leq rank\: R,
\end{equation}
(recall that $R$ is a free $\Cset[\d]$-module), and
\begin{equation}\lbb{surcot}
[T,\ov{\L}_i]=\ov{\L}_{i-1} \quad (i\geq 0).
\end{equation}

\begin{proposition}\lbb{pcannia}
The annihilation algebra $\A(R)$ of a finite simple Lie conformal
superalgebra $R$ is isomorphic as a topological Lie superalgebra to
one of the following linearly compact Lie superalgebras:
\begin{enumerate}
\item $\ov{W}(1,N)$, $N \geq 0$; $\ov{S}(1,N)'$, $N \geq 2$; $\ov{K}(1,N)$,
$N \geq 0$, $N\neq 4$; $C\ov{K}(1,4)$; $\ov{E}(1,6)$;
\item
$\ss[[t]]$, where $\ss$ is a simple finite-dimensional Lie
superalgebra.
\end{enumerate}
\end{proposition}

\begin{proof}
By Lemma \ref{annauc}, $\A(R)$ is an irreducible central extension of
the Lie superalgebra $\Cset[[t_1,\ldots,t_r]]\what{\tt} \ss$, where
$r=0 \;\text{or}\; 1$ and $\ss$ is a simple linearly compact Lie
superalgebra.  It follows from (\ref{ficanna}) and (\ref{dicoquofi})
that $\ell_j:=\ov{\L}_{j+d}$ $(j \in -d + \Zset_+)$ is an algebra
filtration of $\A(R)$ such that
\begin{equation}
dim\; \ell_j/\ell_{j+1}\leq \;rank\: R, \quad j \in \Zset_+,
\end{equation}
hence the growth of this filtration is at most $1$. This filtration
induces an algebra filtration on $1\tt \ss$, whose growth is therefore
at most $1$.  We conclude (see Theorem \ref{grsilc}) that $gw(\ss)
\leq 1$ and therefore, by Theorem \ref{gr1lcls}, either $dim\; \ss
<\infty$ or $\ss$ is one of the Lie superalgebras $\ov{W}(1,N)$ $(N
\geq 0)$, $\ov{S}(1,N)'$ $(N \geq 2)$, $\ov{K}(1,N)$ $(N \geq 0)$,
$\ov{E}(1,6)$.

If $r=0$, then $dim\; \ss=\infty$, since $dim\; \A(R)=\infty$. In this
case, due to Corollary \ref{canniauc}, $\A(R)$ is the universal
central extension of $\ss$. Consequently, by the results of Section
\ref{subsomeex}, $\A(R)$ is one of the Lie superalgebras
$\ov{W}(1,N)$, $\ov{S}(1,N)'$, $\ov{K}(1,N)$, $ C\ov{K}(1,4)$, $
\ov{E}(1,6)$.

If $r=1$ and $dim\; \ss <\infty$, then, by Lemma \ref{annauc}, $\A(R)$
is an irreducible central extension of $\ss[[t]]$. This gives us an
embedding $Der(\A(R)) \subset Der(\ss[[t]])$.  By Proposition
\ref{cohcurr}, the universal central extension of $\ss[[t]]$ is
$\what{\ss}[[t]]$, where $\what{\ss}$ is the universal central
extension of $\ss$. Hence we have a surjective homomorphism
$\varphi_1:\what{\ss}[[t]] \to \A(R)$ such that $ker\: \varphi_1$ is a
central ideal. Also, since any derivation of a Lie superalgebra lifts
uniquely to the universal central extension, we get an embedding (see
Proposition \ref{detp})
\begin{equation*}
Der(\A(R)) \subset Der(\what{\ss}[[t]])=
Der(\Cset[[t]]) \tt 1 + \Cset[[t]] \tt Der(\what{\ss}).
\end{equation*}
Thus, the derivation $T$ of $\A(R)$ induces a derivation of
$\what{\ss}[[t]]$, which we denote by $\what{T}$.  We have:
$\what{T}=P(t)\frac{\d}{\d t}+T_1$, where $P(t) \in \Cset[[t]]$, $T_1
\in \Cset[[t]] \tt Der(\what{\ss})$.

Define a filtration on $L=\Cset \what{T} \ltimes \what{\ss}[[t]]$ by
letting $deg\; t=-deg\; \frac{\d}{\d t}=1$, $deg
\;\what{\ss}=0$: $L \supset L_0 \supset L_1 \supset \ldots$.
Then $\what{T}(L_0) \not \subset L_0$, otherwise $\what{T}$, being
surjective on $\A(R)$, is surjective on $L_0$, hence on $L_0/L_1
\supset \what{\ss}$, which is impossible by Proposition
\ref{dernonsofd} since $\what{\ss}$ is not solvable.  Hence we may
assume that $\what{T} =\frac{\d}{\d t}+T_0$, where $T_0 \in
L_0$. Applying Proposition \ref{derconj} to $L$, $V=0$,
$D=\frac{\d}{\d t}$ and $g_0=T_0$, we may find a continuous
automorphism $\psi$ of $\what{\ss}[[t]]$ that transforms $\what{T}$ to
$\frac{\d}{\d t}$. But $\Cset\frac{\d}{\d t} \ltimes
\what{\ss}[[t]]$ is the extended annihilation algebra of the Lie
conformal superalgebra $Cur \;\what{\ss}$ and the homomorphism
$\varphi_1 \circ \psi$ extends to a surjective homomorphism of extended
annihilation algebras
\begin{equation*}
\varphi \; : \frac{\d}{\d t} \ltimes  \what{\ss}[[t]] \to T \ltimes \A(R)
\end{equation*}
satisfying the conditions of Proposition \ref{dkrecthm}. Hence
$\varphi$ is induced by a surjective homomorphism $\wti{\varphi}\; :
Cur\; \what{\ss} \to R$. But $R$ is simple, hence $\wti{\varphi}$
induces an isomorphism $Cur\; \ss \to R$.

It remains to consider the case $\A(R)/Z\simeq
\Cset[[t]]\what{\tt}\ss$, where $dim\; \ss =\infty$ and $Z$ is a
central ideal. Recall that we may assume, from the proof of Lemma
\ref{annauc} that
\begin{equation}\lbb{exprft}
T=\frac{\d}{\d t} \tt 1 + \sum_i f_i \tt D_i, \quad D_i \in Der(\ss),
\;\;f_i \in \Cset[[t]].
\end{equation}
Note that $L_i=\ov{\L}_{(i+1)d}$ $(i \geq 0)$ is an algebra filtration
of the extended annihilation algebra $\A(R)^e(=L_{-1})$. Denote again
by $\{L_i\}_{i \geq-1}$ the induced filtration on $\A(R)^e/Z\simeq
\Cset T \ltimes (\Cset[[t]]\what{\tt} \ss)$. Let $1\tt \ss_0=L_0 \cap
(1\tt \ss)$ and consider the canonical filtration of $\ss$ associated with
the subalgebra $\ss_0$
\begin{equation*}
\ss=\ss_{-1} \supset \ss_0 \supset \ss_1 \supset \ldots .
\end{equation*}
Consider the following filtration of $\Cset T \ltimes
(\Cset[[t]]\what{\tt}
\ss)=:\wti{L}_{-1}$
\begin{equation*}
\wti{L}_m =\sum_{\substack{i\geq 0,\; j\geq -1  \\ i+j=m }}
t^i\Cset[[t]]\what{\tt}\ss_j
\quad (m \geq 0).
\end{equation*}
Since $\{ \ss_j \}_{j\geq -1}$ is the canonical filtration of $\ss$
associated to $\ss_0$ and $T$ has the form (\ref{exprft}), it is easy
to see that $\{ \wti{L}_m \}_{m \geq -1}$ is the canonical filtration
of $\A(R)^e/Z$ associated to $\wti{L}_0$. By Chevalley's principle,
$\wti{L}_0 \supset L_N$ for some $N >0$, and since $\{\wti{L}_m\}$ is
a canonical filtration, we conclude that $\wti{L}_j \supset L_{N+j}$
for all $j \geq 0$. It follows that
\begin{equation*}
gw(\{\wti{L}_j \}) \leq gw(\{L_j \}) \leq 1
\end{equation*}
(the last inequality follows from (\ref{dicoquofi})). But
$gw(\{\wti{L}_j\})=gw(\{t^j\Cset[[t]]\}) +gw(\ss) =2$. Thus, the
remaining case is impossible.
\end{proof}

\subsection{Derivations of the annihilation algebra}\lbb{dercannia}

\begin{proposition}\lbb{crlcs}
Let $L$ be a linearly compact Lie superalgebra, and let $\ss$ be a
reductive finite-dimensional Lie subalgebra of $L$ such that
$L=\prod_i V_i$, where the $V_i$'s are finite-dimensional irreducible
$\ss$-modules. Then there exists a closed $\ss$-submodule $V$ of
$Der(L)$, complementary to the space of inner derivations $ad(L)$; one
has $[v, ad(a)]=0$ for all $v \in V$ and $a \in \ss$.
\end{proposition}
\begin{proof}
We have: $Der(L) \subset Hom(\oplus _i V_i, \prod_j V_j)=\prod_{i,j}
Hom(V_i, V_j)$, hence $Der(L)$ is an $\ss$-invariant closed (hence
linearly compact) subspace of a direct product of finite dimensional
irreducible $\ss$-modules.  The subspace $ad(L)$ of $Der(L)$ is
$\ss$-invariant and closed too (by Proposition \ref{plcvs} (1)).
Hence there exists a closed $\ss$-invariant complementary subspace
$V$.  But $[ad(\ss), D]\subset ad(L)$ for any $D \in Der(L)$, hence
$[ad(\ss), D]=0$ if $D \in V$.
\end{proof}

We will be working with the following \emph{standard}
$\Zset$-gradation of the Lie superalgebras that occurr in Proposition
\ref{pcannia} (1):
\begin{alignat}{3}\lbb{stgrlc}
&\ov{W}(1,N)=\prod_{j\geq -1} W(1,N)_j; \;\;&
&\ov{S}(1,N)'=\prod_{j\geq -1}S(1,N)'_j; \;\;&
&\ov{S}(1,2)'=\prod_{j\geq -2}S(1,2)^0_j;  \\ \nonumber
&\ov{K}(1,N)=\prod_{j\geq -2} K(1,N)_j; \;\;&
&C\ov{K}(1,4)=\prod_{j\geq -2} CK(1,4)_j; \;\;&
&\ov{E}(1,6)=\prod_{j\geq -2} E(1,6)_j;
\end{alignat}
here the gradations of depth $1$ (resp. $2$) are defined by letting
\begin{equation*}
deg\; x=- deg\; \d_0=1 \;(\text{resp.}\; 2),
\quad deg \; \xi_i =-deg\; \d_i
=1,
\end{equation*}
and in the $C\ov{K}(1,4)$ case we let $deg(\text{center})=0$.

\begin{lemma}[\cite{K5}]\lbb{derwnkne}
All continuous derivations of $\ov{W}(1,N)$, $\ov{K}(1,N)$,
$C\ov{K}(1,4)$ $\ov{E}(1,6)$ are inner.
\end{lemma}


\begin{proof}
Let $\ss$ be the even part of $W(1,N)_0$, $K(1,N)_0$, $E(1,6)_0$.  We
have $\ss=\Cset \oplus \gl_N$, $\mathfrak c \so_N$, $\mathfrak c\so_6$
respectively and the representation of $\ss$ on $W(1,N)_{-1}$,
$K(1,N)_{-1}$, $E(1,6)_{-1}$ is the direct sum of the standard $\gl_N$
and $\gl_1$- modules, the standard $\mathfrak c \so_N$- and $\mathfrak
c \so_6$-module respectively.  By Proposition \ref{crlcs}, in all
cases, $Der(L)= ad (L) \oplus V$ as $\ss$-module, and any $D \in V$ is
an $\ss$-module homomorphism.

We remark that $x \d_0 + \sum_i \xi_i \d_i \in \ss$ for $\ov{W}(1,N)$
and $2x \d_0 + \sum_i \xi_i \d_i \in \ss$ for $\ov{K}(1,N)$,
$C\ov{K}(1,4)$ and $\ov{E}(1,6)$. It follows that any $D \in V$
preserves the standard gradation of $L$.

By Schur Lemma $D=diag(\lambda, \mu,\ldots, \mu)$ on $W(1,N)_{-1}$.
It follows that the grading preserving derivation $D'=D - ad (\lambda
x\d_0 + \mu \sum_i \xi_i \d_i)$ is zero on $W(1,N)_{-1}$ and it is an
$\ss$-module homomorphism.

Let $y \in W(1,N)_k$, and $g_{-1} \in W(1,N)_{-1}$. By induction on $k
\geq -1$ we have $0=D'([y,g_{-1}])=[D'y, g_{-1}]$. Hence by
transitivity we conclude that $D'y=0$, i.e.  $D'=0$ on
$W(1,N)_k$. Consequently $D'=0$ on $\ov{W}(1,N)$ and
$D=ad(\la x\d_0 +\mu\sum_i \xi_i \d_i )$.
On the other hand, $D \in V$, hence $\la=\mu=0$.
Therefore $D=0$ and every derivation of $\ov{W}(1,N)$  is inner.

A similar method can be used in the remaining cases.  Notice that we
may exclude the case $\ov{K}(1,2)$, which is isomorphic to
$\ov{W}(1,1)$, so that $K(1,N)_{-1}$ (resp. $E(1,6)_{-1}$) is
irreducible as a $\so_N$-module (resp. $\so_6$-module), hence any
derivation $D \in V$ acts on this subspace as a scalar matrix. Now we
proceed as above using also that in both cases the $-2^{\text{nd}}$
component is the bracket of the $-1^{\text{st}}$ component with
itself. The result for $C\ov{K}(1,4)$ easily follows once one has
established it for $\ov{K}(1,4)$.
\end{proof}

In the next two lemmas we will use the following outer derivations of
$\ov{S}(1,N)'$ $( \subset \ov{W}(1,N))$: $E=ad(\xi_1\ldots \xi_N
\d_0)$, $H=ad(\sum_{i=1}^N \xi_i \d_i)$.

\begin{lemma}[\cite{K5}]\lbb{dersnnbig}

$Der(\ov{S}(1,N)')= ad (\ov{S}(1,N)') \oplus \Cset E \oplus \Cset H$
for $ N >2$.
\end{lemma}


\begin{proof}
The Lie algebra $\gl_N$ is the even part of $S(1,N)'_0$. By Lemma
\ref{crlcs}, we have $Der(\ov{S}(1,N)') =ad (\ov{S}(1,N)') \oplus V$
as $\gl_N$-modules and any $D \in V$ is a $\gl_N$-module
homomorphism.  Let us denote by $(R(\pi), m)$ the irreducible
$\sl_N$-module $R(\pi)$ whose eigenvalue with respect to the operator
$Nad(x \d_0) + H$ is $m$.  The irreducible $\gl_N$-modules
which appear more than once in the standard gradation (\ref{stgrlc})
are: $(R(\pi_{N-1}), -1)$, which occurs in $S(1,N)'_{-1}$ and
$S(1,N)'_{N-2}$ and $(R(\pi_{N-1}), N-1)$ which occurs in
$S(1,N)'_{0}$ and $S(1,N)'_{N-1}$.

Let $v_1=\d_N$ (resp. $v_2=\xi_1 \ldots \xi_{N-1} \d_0$) be the
highest weight vector of the module $(R(\pi_{N-1}), -1)$ in
$S(1,N)'_{-1}$ (resp. in $S(1,N)'_{N-2}$). Then $[\xi_1 \ldots \xi_N
\d_0,\d_N]= \xi_1 \ldots \xi_{N-1} \d_0$. Now,
$D$ is a $\gl_N$-module homomorphism, hence $D(v_1)=\al v_1 +\beta
v_2$ and $D - \beta E$ maps $S(1,N)'_{-1}$ into itself.  Also,
$S(1,N)'_{-1}$ is sum of two non-isomorphic, irreducible
$\sl_N$-modules, so by Schur Lemma we have that $D - \beta E=diag(\lambda,
\mu,\ldots, \mu)$.
Consequently, $D'= D - \beta E -ad (\lambda x\d_0 +
\mu \sum_i \xi_i \d_i)$ acts
as $0$ on $S(1,N)'_{-1}$. Suppose $y \in S(1,N)'_{k}$ and $g_{-1} \in
S(1,N)'_{-1}$. By induction on $k \geq 1$, we have
$0=D'([y,g_{-1}])=[D'y,g_{-1}]$. Note that the homogeneous components
of $D'y$ have degree greater or equal than $0$. By transitivity, we
conclude that $D'y=0$, hence $D'=0$ on $\ov{S}(1,N)'$, and therefore
$D$ can be expressed as a linear combination of $E$, $ad(x\d_0)+H$ and
some inner derivation. Since $ad(Nx\d_0) + H$ is an inner derivation,
the lemma is proved.
\end{proof}


\begin{lemma}[\cite{K5}]\lbb{ders2}
$Der(\ov{S}(1,2)')=ad(\ov{S}(1,2)') \oplus \sl_2$, where the standard
basis of $\sl_2$ consists of $E$, $H$ as above and $F$, defined as
follows:
\begin{equation*}
F ( P(x) \xi_2 \d_0 - \d_0 P(x) \xi_1 \xi_2 \d_1) = -P(x) \d_1, \quad
F ( P(x) \xi_1 \d_0 + \d_0 P(x)\xi_1 \xi_2 \d_2), = P(x)\d_2,
\end{equation*}
$F(P(x)S(2))=0$ and $F(P(x)\d_0- 1/2\d_0 P(x)(\xi_1\d_1 +
\xi_2\d_2))=0$.
\end{lemma}

\begin{proof}
With respect to the action of $H$, $\ov{S}(1,2)'$ decomposes into
eigenspaces relative to the eigenvalues $\{-1,0,1 \}$. The even part
of $\ov{S}(1,2)'$ is contained in the zero eigenspace. Also, $E$
(resp. $F$) transforms the $-1$ (resp. $+1$) eigenspace into the $+1$
(resp. $-1$) eigenspace and kills the other eigenspaces.  We will use
the standard depth 2 gradation of $\ov{S}(1,2)'$, see
(\ref{stgrlc}). The even part of $\ov{S}(1,2)^0_0$ is $\gl_2$.  Lemma
\ref{crlcs} implies that $Der(\ov{S}(1,2)')=ad(\ov{S}(1,2)') \oplus V$
as $\gl_2$-modules and any $D
\in V$ is a $\gl_2$-module homomorphism.
The modules occurring more than once are $(R(\pi_1), -1)$ (in
$S(1,2)^0_{-1}$ and $S(1,2)^0_{0}$), and $(R(\pi_1), 1)$ (in
$S(1,2)^0_{0}$ and $S(1,2)^0_{1}$).

Let $v_1=\d_2$ (resp. $v_2= \xi_1 \d_0$) be the highest weight vector
of $(R(\pi_1), -1)$ in $S(1,2)^0_{-1}$ (resp. $S(1,2)^0_{0}$). Then
$E(v_1)=v_2$.  The derivation $D$ is a $\gl_2$-module homomorphism, so
$D(v_1)= \alpha v_1 + \beta v_2$, and $D - \beta E$ maps
$S(1,2)^0_{-1}$ into itself. By Schur Lemma, $D -
\beta E=diag(\lambda,\mu,\mu)$. It follows that  the derivation
$D'= D - \beta E -ad(\lambda x \d_0 +
\mu \sum_i \xi_i \d_i)$
acts as $0$ on $S(1,2)^0_{-1}$.

Let $y \in S(1,2)^0_{0}$. We have
\begin{equation*}
S(1,2)^0_{0} =(R(\pi_1), -1) \oplus \left( (R(2 \pi_1), 0) \oplus
(R(0), 0) \right) \oplus (R(\pi_1), 1),
\end{equation*}
so that $y= y_{-1} + y_0 + y_{1}$. For any $g_{-1} \in S(1,2)^0_{-1}$,
$0=D'([y, g_{-1}])= [D'y_{1},g_{-1}]$, hence $[D'y,g_{-1}] +
[D'y_{0},g_{-1}] + [D'y_{1},g_{-1}]=0$.  Now, $(R(2 \pi_1), 0)$ and
$(R(0), 0)$ occur only in $S(1,2)^0_{0}$, hence $D'y_{0} \in
S(1,2)^0_{0}$. The module $(R(\pi_1), 1)$ occurs in $S(1,2)^0_{0}$ and
$S(1,2)^0_{1}$, so $D'y_{1} \in S(1,2)^0_{0} \oplus S(1,2)^0_{1}$.
Also, $D'y_{-1} \in S(1,2)^0_{-1} \oplus S(1,2)^0_{0}$.  We may assume
that $y_{-1}=\xi_1 \d_0$. Then $D'y_{-1}=\al
\d_2 + \beta \xi_1 \d_0$.  $F$ kills $((R(2 \pi_1), 0) \oplus (R(0), 0) )
\oplus (R(\pi_1), 1)$ and $F(\xi_1 \d_0)=\d_2$.
It follows that $(D'-\al F)(\xi_1 \d_0) =\beta \xi_1
\d_0 \in S(1,2)^0_{0}$.  Also, $D''=D'-\al F$ is still
identically zero on $S(1,2)^0_{-1}$.  Hence
we have $0=D''([y,g_{-1}])=[D''y,g_{-1}]$, and $D''y$ has homogeneous
components of degree greater or equal to zero. Transitivity implies
that $D''=0$ on $S(1,2)^0_{0}$. Induction on $k \geq1$ shows that
$D''=0$ on $S(1,2)^0_{k}$. Therefore,
D can be written as a linear combination of $E$, $H$, $F$ and some
inner derivation.
\end{proof}
\vspace{1mm}

\subsection{The classification theorem}\lbb{clthmfsliecs}

\begin{lemma}\lbb{l0ntinv}
Let $L=\prod_{j\geq -d}\: \g_j$ be one of the linearly compact Lie
superalgebras that occur in Proposition \ref{pcannia} (1) with the
standard gradation (\ref{stgrlc}) and let $L_0=\prod_{j \geq 0}\g_j$.
Let $T$ be an even surjective derivation of $L$. Then $T(L_0) \not
\subset L_0$.
\end{lemma}

\begin{proof}
In the contrary case, we also have $T(L_1)\subset L_1$ and therefore
$T$ induces a surjective derivation of $\g_0 \simeq L_0/L_1$. But for
$\ov{W}(1,N)$, $\ov{S}(1,N)'$, $\ov{K}(1,N)$ and $\ov{E}(1,6)$ we have
$\g_0 \simeq \gl(1,N)$, $\sl(1,N)$, $\mathfrak c \so_N$, $\mathfrak c
\so_6$ respectively. Hence by Proposition \ref{dernonsofd} we reach a
contradiction, unless $L=\ov{W}(1,0)$, $\ov{W}(1,1)\simeq\ov{K}(1,2)$
or $\ov{K}(1,1)$. The second case is also excluded since $\gl(1,1)$
has only inner derivations (this is immediate by Proposition \ref{crlcs})
and the third case is reduced to the first
one since the even part of $\ov{K}(1,1)$ is $\ov{W}(1,0)$. But if
$T=\sum_{j\geq0} a_j x^j\d_0 \in\ov{W}(1,0)$, it is easy to
see that one of the elements $\d_0$ or $x\d_0$
does not lie in the image of $ad(T)$.
\end{proof}

\begin{theorem}\lbb{pcanniaader}
Let $R$ be a finite simple Lie conformal superalgebra. The following
list gives all the linearly compact Lie superalgebras that can occur
as extended annihilation algebras $\A(R)^e$:
\begin{enumerate}
\item $\Cset ad( \d_0) \ltimes \ov{W}(1,N),\; N\geq 0$;
\item $(\Cset ad( \d_0 - a \sum_{i=1}^N \xi_i \d_i))
      \ltimes \ov{S}(1,N)'$,  $N \geq  2$;
\item $(\Cset ad( \d_0 - \xi_1 \ldots \xi_N \d_0) \ltimes \ov{S}(1,N)'$,
$N$ even, $N \geq 2$;
\item $\Cset ad(\d_0) \ltimes \ov{K}(1,N),\;N\geq 0, \; N \neq 4$;
\item $\Cset ad(\d_0) \ltimes C\ov{K}(1,4)$;
\item $\Cset ad(\d_0) \ltimes \ov{E}(1,6)$;
\item $\Cset \frac{\d}{\d t} \ltimes \ss[[t]]$,
      where $\ss$ is a finite-dimensional simple Lie superalgebra.
\end{enumerate}
\end{theorem}
\begin{proof}
Recall that $\A(R)^e=\Cset T \ltimes L$, where $L=\A(R)$ is one of the
linearly compact Lie superalgebras listed in Proposition \ref{pcannia}
and $T$ is an even surjective derivation of $L$.

If $L$ is one of the Lie superalgebras listed in Proposition
\ref{pcannia} (1),
consider the filtration $\{L_i\}$ of $L$ corresponding to the
standard gradation $\prod_{j \geq -d}\: \g_j$  of $L$ (cf. \ref{stgrlc}).
One checks directly that
in all cases one has:
\begin{equation}\lbb{d0surj}
[\d_0, \g_j]=\g_{j-d}\quad \text{for all} \quad j \geq d.
\end{equation}

Furthermore, due to lemmas \ref{derwnkne}-\ref{ders2} and
\ref{l0ntinv} we have
\begin{equation}\lbb{genexprft}
T=c\:ad(\d_0) +v + ad(g_0),
\end{equation}
where $c \in \Cset$ is non-zero, $g_0 \in L_0$ and $v \in V$, where
$V$ is one of the following subspaces of $Der(L)$:
\begin{align*}
&V=0 \;\;\text{if} \;\;  L=\ov{W}(1,N),\; \ov{K}(1,N), \;C\ov{K}(1,4)
\;\; \text{or} \;\; \ov{E}(1,6);   \\
&V=\Cset H \;\; \text{if} \;\; L=\ov{S}(1,N)', \;\; N \;\; \text{odd}
\;\; \text{(cf. Lemma \ref{dersnnbig})};   \\
&V=\Cset E \oplus \Cset H \;\; \text{if} \;\; L=\ov{S}(1,N)',
\;\;N \;\;\text{even},\;\;  N>2 \;\;
\text{(cf. Lemma \ref{dersnnbig})};     \\
&V=\Cset E \oplus \Cset H \oplus \Cset F \;\; \text{if}
\;\;L=\ov{S}(1,2)'\;\;\text{ (cf. Lemma \ref{ders2})}.
\end{align*}

We may apply now Proposition \ref{derconj} to $D=c\: \d_0$ (cf.
(\ref{genexprft})) and the above $V$ since (\ref{derfilpr}) holds due
to (\ref{d0surj}) and (\ref{outderfilpr}) also obviously holds. Hence
by an inner automorphism of $L$ we can bring $T$ to the form:
\begin{equation*}
T=c\: ad(\d_0) +v, \quad \text{where} \quad c \in
\Cset\backslash \{0 \}, \;\;v \in V.
\end{equation*}
By rescaling we can make $c=1$ and, using an inner automorphism of the
Lie algebra $V$, $T$ can be brought further, in all $\ov{S}(1,N)'$
cases, to the form $ad(\d_0)-E$ (if $N$ is even) or $ad(\d_0) -aH$, $a
\in \Cset$.

The case of $L=\ss[[t]]$ has been treated in a similar fashion in the
proof of Proposition \ref{pcannia}.

\end{proof}
\begin{theorem}\lbb{clthm}
Any finite simple Lie conformal superalgebra is isomorphic to one of
the Lie conformal superalgebras of the following list:
\begin{enumerate}
\item $W_N$, $N \geq 0$;
\item $S_{N,a}$, $N \geq 2$, $a \in \Cset$;
\item $\wti{S}_N$,  $N$ even, $N \geq 2$;
\item $K_N$, $N \geq 0$, $N\neq 4$;
\item $K_4'$;
\item $CK_6$;
\item $Cur\; \ss$, where $\ss$ is a simple finite-dimensional Lie
superalgebra.
\end{enumerate}
\end{theorem}

\begin{proof}
It follows from Theorem \ref{pcanniaader} and Proposition
\ref{dkrecthm}.
\end{proof}

A formal distribution Lie superalgebra $(\g,\F)$ is called
\emph{simple} if it contains no non-trivial regular ideals; it is
called \emph{finite} if the $\Cset[\d_z]$-module $\ov{\F}$ is finitely
generated. Two formal distribution Lie superalgebras $(\g,\F)$ and
$(\g_1,\F_1)$ are called \emph{isomorphic} if there exists an
isomorphism $\varphi : \g \to \g_1$ such that
$\varphi(\ov{\F})=\ov{\F_1}$.

The correspondence between Lie conformal superalgebras and formal
distribution Lie superalgebras implies the following corollary of
Theorem \ref{clthm}.

\begin{corollary}[\cite{K7}]\lbb{clthmfdls}
A complete list of finite simple formal distribution Lie superalgebras
consists of quotients of loop algebras $(\ss[t,t^{-1}]/(P),
\F_{\ss})$, where $P$ is a non-invertible polynomial of
$\Cset[t,t^{-1}]$ and $\ss$ is a simple finite-dimensional Lie
superalgebra, $(W((1,N)),\F_W)$ $(N \geq 0)$, $(S((1,N,a))',
\F_{S,a})$ $(N \geq 2,\;a \in \Cset)$, $(\wti{S}((1,N)),
\F_{\wti{S}})$ $(N \geq 2, \;\; N \text{even})$, $(K((1,N)), \F_K)$
$(N\geq 0,\; N \neq 4)$,
\newline
$(CK((1,4))', \F_K)$,
$(K((1,4))', \F_{K'})$,
$(CK((1,6)), \F_{CK})$.
\end{corollary}


\end{document}